\documentclass[twocolumn, journal]{IEEEtran}
\IEEEoverridecommandlockouts

\usepackage{color}
\usepackage{bm}
\usepackage{graphicx}
\usepackage{amsmath}
\usepackage{amssymb}
\usepackage{algorithm}
\usepackage{algorithmic}
\usepackage{multirow}
\usepackage{booktabs}
\usepackage{array}
\usepackage{lipsum}
\usepackage{enumerate}
\usepackage{stfloats}
\usepackage{subfigure}
\usepackage{cases}
\usepackage{diagbox}
\usepackage{cite}
\usepackage{balance}

\usepackage{enumitem}

\newtheorem{defi}{Definition}

\newtheorem{prop}{Proposition}

\allowdisplaybreaks[4]
\ifCLASSINFOpdf
\else
\fi
\makeatletter

\newcommand{\Rmnum}[1]{\expandafter\@slowromancap\romannumeral #1@}
\makeatother

\begin{document}
% not capitalized unless they are the first or last word of the title..
%\title{Spectrally Compatible Waveform Design for MIMO Radar Multi-Target Detection in the Presence of Signal-Dependent Interference}
%\title{\color{red}Hybrid Beamforming Design   for  Dual-Functional MIMO Radar-Communication System Using Multi-Frequency Signalings}
%\title{Hybrid Beamforming and DOA  Estimation for QoS-Constrained  Multi-Carrier Dual-Function Radar-Communication System}
\title{ {Task-Oriented Hybrid Beamforming for OFDM-DFRC Systems with Flexibly Controlled Space-Frequency Spectra}}

\author{\IEEEauthorblockN{Lingyun Xu, \IEEEmembership{Student Member, IEEE,}
Bowen Wang,~\IEEEmembership{Student Member, IEEE,}
and Ziyang Cheng,~\IEEEmembership{Member, IEEE}% <-this %
\vspace{-2em}
}\\  
% \thanks{L. Xu is with the School of Information and Communication Engineering, University of Electronic Science and Technology of China, Chengdu, China, and with the School of Electronics and Electrical Engineering, University of Glasgow, Scotland, UK. (email: xusherly@163.com).}
\thanks{L. Xu, B. Wang, and Z. Cheng are with the School of Information and Communication Engineering, University of Electronic Science and Technology of China, Chengdu, China. (email: xusherly@163.com, bwwang@std.uestc.edu.cn, zycheng@uestc.edu.cn).}% <-this % 
}

\maketitle

%{Shell \MakeLowercase{\textit{et al.}}: Bare Demo of IEEEtran.cls for IEEE Transactions on Magnetics Journals}



% for Transactions on Magnetics papers, we must declare the abstract and
% index terms PRIOR to the title within the \IEEEtitleabstractindextext
% IEEEtran command as these need to go into the title area created by
%\IEEEtitleabstractindextext{%
\begin{abstract}
% 1. Our Goal----------------
This paper investigates the issues of the hybrid beamforming design for the orthogonal frequency division multiplexing dual-function radar-communication (DFRC) system in multiple task scenarios involving the radar scanning and detection task and the target tracking task.
% 2. Our contributions (Proposed XX Problem)----------------
To meet different task requirements of the DFRC system, we introduce two novel radar beampattern metrics, the average integrated sidelobe to minimum mainlobe ratio (AISMMR) and average peak sidelobe to integrated mainlobe ratio (APSIMR), to characterize the space-frequency spectra in different scenarios.
Then, two HBF design problems are formulated for two task scenarios by minimizing the AISMMR and APSIMR respectively subject to the constraints of communication quality-of-service (QoS), power budget, and hardware.
% 3. Our Methods----------------
Due to the non-linearity and close coupling between the analog and digital beamformers in both the objective functions and QoS constraint, the resultant formulated problems are challenging to solve.
 {Towards that end, a unified optimization algorithm based on a consensus alternating direction method of multipliers (CADMM) is proposed to solve these two problems.
Moreover, under the unified CADMM framework, the closed-form solutions of primal variables in the original two problems are obtained with low complexity.}
% 4. Our Results----------------
Numerical simulations
are provided to demonstrate the feasibility and effectiveness of
the proposed algorithm.

\end{abstract}


\IEEEpeerreviewmaketitle

\vspace{-0.5em}
\section{Introduction}

\IEEEPARstart{T}{he} proliferation of wireless mobile services exhibits an exponential trend, leading to a scarcity of spectral resources and to escalating spectrum prices.
Even worse, radar competes for the limited spectrum adding to the spectrum shortage problem and leading to mutual interference between communication and radar systems \cite{liu2022integrated,zhang2021overview,hassanien2019dual}. 
For example, the spectrum of long-term evolution (LTE) communication systems is partially overlapped with S-band radar systems  \cite{6817773}, and the spectrum of global system for mobile (GSM) communication systems overlaps with ultra-high frequency (UHF) radar systems \cite{7470514}. 
To address the aforementioned radar and communication spectrum congestion, the research of radar and communication spectrum sharing (RCSS) technology \cite{liu2022integrated,zhang2021overview,hassanien2019dual} is well-underway.
Generally, there are two directions to realize RCSS: 1) radar-communication co-existence (RCC), and 2) dual-function radar-communication (DFRC).

For the RCC technology, the radar and communication functionalities are operated in separate devices and share the same spectrum.
 {The early works on RCC were based on cognitive radio (CR) technology \cite{8168273,CR2008Wang,8253420}, where one system could access the spectrum allocated to the other system based on the spectrum sensing capability of CR.
For example, dynamic spectrum access was proposed in \cite{8168273}, where the radar transmitted and received in the available disjoint narrow bands by sensing the spectrum at low sampling and processing rates.
Then, opportunistic spectrum access to radar bands was studied in \cite{CR2008Wang}, where the spatial and temporal opportunities in the radar spectrum were explored by CR and the throughput of communication was increased by increasing the tolerance to radar interference.}
The above works can implement the RCC, however, it does not really achieve the RCSS since one system only has restricted bandwidth and time to access to the spectrum assigned to the other system, leading to its limited performance.
 As a further step, some researchers have proposed the joint design of radar and communication systems to realize RCC \cite{JRCsurvey2020Feng,8335405,8048004,7953658,8233171,Joint2017Her,He2022fdCRC}. 
 For example, the radar space-time transmit code, radar receive filter, and communication space-time transmit covariance matrix were jointly optimized in \cite{8335405}.
Then, the authors in \cite{8048004} jointly designed the radar waveform and communication symbol encoding matrix.
Besides, the authors in \cite{7953658} jointly designed the radar subsampling scheme and the communication transmit covariance matrix, where the signal-to-interference plus noise ratio (SINR) of radar systems was maximized while guaranteeing the communication rate requirements.
Although the joint design approach can achieve satisfactory radar and communication performance, it has two limitations: 1) The radar and communication tasks are implemented in two devices, which leads to extra hardware costs. 2) The joint design approach requires frequent information exchange and interference management, which is hard to realize in practice.
Therefore, the DFRC is proposed to overcome the above limitations.

For the DFRC, the radar and communication functionalities are integrated into one common platform, operating the same signal processing framework, sharing the same frequency band, and reducing hardware costs.
Due to the numerous advantages offered by DFRC, it is envisioned to be a promising technique in beyond 5G (B5G) and 6G network \cite{liu2022integrated,zhang2021overview,hassanien2019dual}.
Prior works realize the DFRC by modifying existing waveforms and protocols.
For example, the authors in \cite{JointBekar2021} proposed to embed the phase shift key (PSK) communication symbols into radar linear frequency modulation (LFM) waveform.
Following this work, to increase the communication rate, the authors in \cite{FHDFRC2022Wu} considered embedding PSK symbols into radar frequency-hopping waveform.
Unlike \cite{JointBekar2021} and \cite{FHDFRC2022Wu}, \textit{Preeti Kumari et al.} realized DFRC by modifying the IEEE 802.1ad standard \cite{8917703}, where the preamble was utilized to enhance radar velocity estimation accuracy.
The aforementioned works \cite{FHDFRC2022Wu,JointBekar2021,8917703} can easily implement the DFRC in practice, however, they highly rely on the existing waveforms and protocols, which limits the design degree of freedom and causes performance loss.
Then, the researchers turn their interests to optimizing DFRC waveforms \cite{Liu2018optwaveform,Liu2020MUEDFRC,Liu2022cramerRC}, where the DFRC waveforms are directly designed by considering radar and communication requirements.
 For example, the authors in \cite{Liu2018optwaveform} proposed a DFRC waveform considering both omnidirectional and directional scenarios, where a low-complexity algorithm was devised to tradeoff radar and communication performance.
In \cite{Liu2020MUEDFRC}, a joint transmit beamforming design for DFRC systems was proposed, where the radar probing signal was optimized with the SINR requirement.
Moreover, the DFRC system design with computation was proposed in \cite{Ding2022fdDFRC}, where both communication symbols and radar waveforms were optimized.
 To increase the communication rate, beamforming design for MIMO orthogonal frequency division multiplexing (OFDM) DFRC systems has also been investigated in \cite{Rahman2020JCRC,Tian2021ofdmdfrc,Liu2022OFDMdfrc,Hu2022OFDMdfrc,John2022OFDMdfrc}.
An OFDM unified system platform for joint communication and radar sensing was proposed in \cite{Rahman2020JCRC}, where downlink and uplink radar sensing functions were integrated into the mobile communication network.
Besides, authors in \cite{Tian2021ofdmdfrc} proposed a transmit/receive beamforming for MIMO-OFDM-DFRC systems with a single communication receiver, where the word error probability for communication and false alarm probability for radar were considered.
Furthermore, authors in \cite{John2022OFDMdfrc} studied an OFDM-DFRC system sensing multiple spatial directions and serving multiple users, where the beamformer and receive filter were optimized subject to error rate and beampattern constraints.

With massive MIMO and millimeter wave (mmWave) becoming promising technologies in the future B5G and 6G networks, mmWave DFRC has received a lot of attention.
However, the aforementioned works \cite{Liu2018optwaveform,Liu2020MUEDFRC,Liu2022cramerRC,Ding2022fdDFRC,Rahman2020JCRC,Tian2021ofdmdfrc,Liu2022OFDMdfrc,Hu2022OFDMdfrc,John2022OFDMdfrc} were based on fully digital architecture, where each antenna is equipped with one unique radio frequency (RF) chain.
Therefore, tremendous power consumption and hardware costs are incurred when extending the works above \cite{Liu2018optwaveform,Liu2020MUEDFRC,Liu2022cramerRC,Ding2022fdDFRC,Rahman2020JCRC,Tian2021ofdmdfrc,Liu2022OFDMdfrc,Hu2022OFDMdfrc,John2022OFDMdfrc} to mmWave DFRC scenarios with massive MIMO.
 To address this drawback, hybrid beamforming (HBF) architecture was proposed in \cite{ahmed2018survey,mishra2019toward,heath2016overview} to save RF chains, in which a small number of RF chains was utilized to implement the digital beamformer and a large number of phase shifters (PSs) to realize the analog beamformer.
The HBF architecture has been widely studied in mmWave communication fields \cite{ahmed2018survey,mishra2019toward,heath2016overview,7913599,8310586,8889665}, and it is then extended to mmWave DFRC system \cite{liu2020joint,9868348,HBFMC2021Cheng,9562280}. 
 Specifically, the authors in \cite{liu2020joint} first proposed a DFRC system with HBF architecture, where the transmit and receive beamforming were designed by considering the tasks of the channel estimation, downlink/uplink communication, and radar target tracking.
Then, the authors in \cite{9868348} investigated HBF design for DFRC systems, where CRB was optimized to improve the direction of arrival (DoA) estimation performance.
Moreover, HBF design for wideband OFDM-DFRC systems was studied in \cite{9562280}, where communication bandwidth efficiency and radar spatial spectrum matching error were jointly optimized based on the weighted optimization criterion.
Following it, the authors in \cite{Liao2023TRofdmdfrc} devised the HBF for OFDM-DFRC systems, where transmit HBF and receive beamformers were jointly designed by maximizing the weighted sum of radar SINR and communication subject to hardware constraints.
Nevertheless, there are three limitations of work \cite{9562280,Liao2023TRofdmdfrc}:
1) Formulation aspect: there was a lack of knowledge of the weights on radar and communication performances, and thus it was difficult to guarantee the performance of both the communication QoS and radar detecting requirement.
2) Waveform aspect: this method just provided a simple way to control the transmit beampattern, but without consideration of the sidelobe levels of the beampattern.
3) Scenario aspect: they only considered a simple scenario with a single DFRC task, so the flexibly controlled space-frequency spectrum required by the multi-task DFRC system cannot be realized.

Motivated by the facts mentioned above, this paper investigates a task-oriented HBF design for wideband OFDM-DFRC systems by flexibly controlling the transmit beampattern.
Specifically, the main contributions of this work can be summarized as follows. 
\begin{itemize}
% \vspace{-0.5em}
\item 
We consider multiple task scenarios in the DFRC system.
The first task is radar scanning and detection, where the level of the wide mainlobe should be maintained to search for the target over a wide area.
The second task is target tracking, where the level of the sidelobe should be suppressed to reduce sidelobe clutters in the presence of strong interference.
For different tasks, we introduce two novel radar beampattern metrics, the average integrated sidelobe to minimum mainlobe ratio (AISMMR) and average peak sidelobe to integrated mainlobe ratio (APSIMR), to flexibly control transmit beampattern.
To achieve multiple tasks while guaranteeing radar space-frequency spectrum behavior and meeting communication QoS, we formulate two problems by minimizing the radar AISMMR and APSIMR respectively subject to data rate, power budget, and analog beamformer constraints.

\item  
We propose a unified algorithm for the multi-task DFRC system to solve the formulated HBF design problems.
Specifically, in order to solve the non-convex optimization problems, we first transform the complicated objective functions into tractable ones and then the primal variables are decoupled using the consensus alternating direction method of multipliers (CADMM) framework.
Under this framework, a unified algorithm is proposed to optimize the hybrid transmit and digital receive beamformers for different task requirements alternately.

\item  Extensive simulation results are illustrated to exhibit the effectiveness of our proposed algorithms.
The results show that our task-oriented HBF design can achieve the desired radar beampatterns for different task scenarios with satisfying communication QoS.
To be more specific, the high mainlobe level is maintained in the scanning and detection task, and the sidelobe level is suppressed in the target tracking task.
To further evaluate the radar sensing performance, we also provide the DOA estimation and radar detection performance of the proposed HBF, verifying the flexible control of the space-frequency spectra in different radar sensing tasks.

\begin{figure*}[!htbp]
	\centering
	\includegraphics[width=0.95\linewidth]{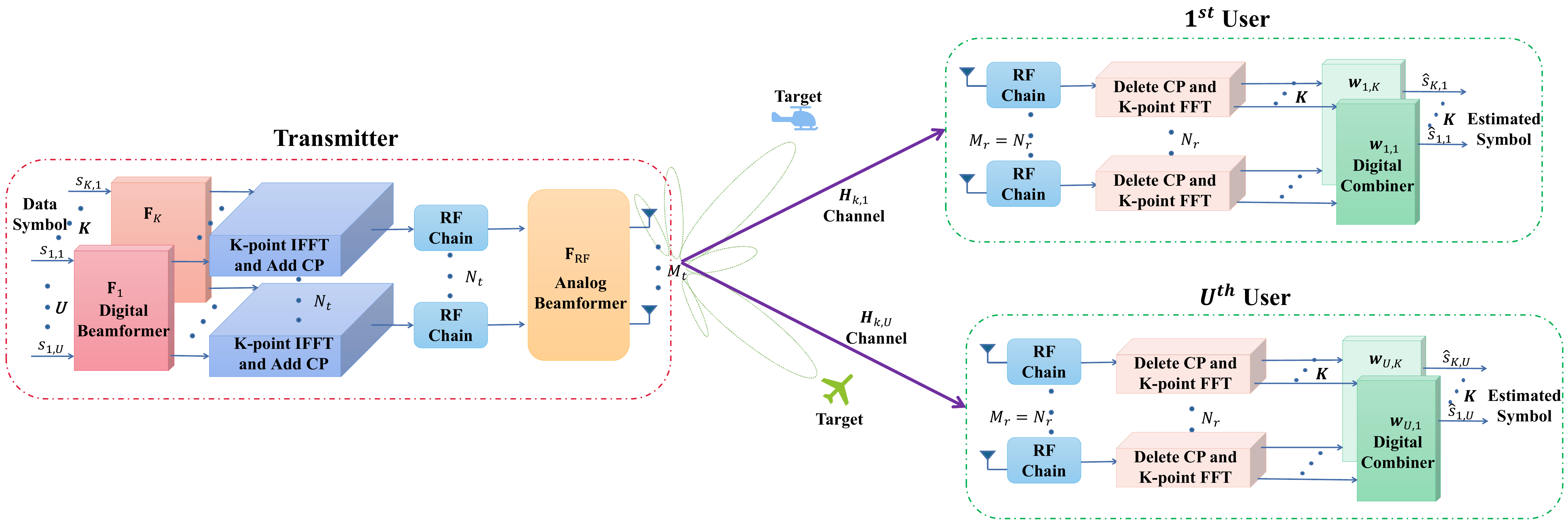}
	\vspace{-1em}
	\caption{A wideband OFDM-DFRC system with a hybrid beamforming structure at the transmitter and a fully digital beamforming structure at the user.}
	% \vspace{-1em}
	\label{fig:pic1}
 \vspace{-1.5em}
\end{figure*}

\end{itemize}

The remainder of this paper is organized as follows.
 {Section II introduces the system model and problem formulation, Section III proposes a unified optimization approach to HBF design for the multi-task DFRC system, Section IV presents simulation results, and conclusions are drawn in Section V.

\textit{Notation:} This paper uses lower-case letters $a$ for scalars, lower-case bold letters $\mathbf{a}$ for vectors and upper-case bold letters $\mathbf{A}$ for matrices. For a matrix $\mathbf{A}$, the element in the $i$-th row and the $j$-th column is denoted by ${\bf{A}}(i,j)$. $\mathbb{C}^{n}$ and $\mathbb{R}^{n}$ denote an $n$ dimensional complex-valued vector and real-valued vector, respectively. $\mathbb{C}^{m\times n}$ and $\mathbb{R}^{m\times n}$ denote an $m$ by $n$ dimensional complex-valued space and real-valued space, respectively. $(\cdot)^T$ and $(\cdot)^H$ denote the transpose and conjugate transpose operators, respectively. $\left|  \cdot  \right|$ represents a determinant or absolute value relying on the context. ${\left\|  \cdot  \right\|_F}$ and ${\rm Tr}({\cdot})$ denote the Frobenius norm and trace, respectively. ${\mathbb E}\left\{   \cdot \right\}$ denotes expectation and $\Re \left\{  \cdot  \right\}$ denotes the real part of a complex-valued number. $\mathcal{C}\mathcal{N}\left( {0,{\text{ }}{\mathbf{R}}} \right)$ denotes the zero-mean complex Gaussian distribution with covariance matrix ${\mathbf{R}}$.

\vspace{-0.5em}
\section{System Model and Problem Formulation}
\vspace{-0.2em}

In this section, we first introduce the system and task scenario, then describe the communication model and the radar model with their performance metrics.
Finally, we formulate the HBF design problems for multiple tasks.

\vspace{-1.2em}
\subsection{ System and Task Scenario}
\vspace{-0.3em}

As shown in Fig.\ref{fig:pic1}, in this paper we consider a wideband OFDM-DFRC system, which sends communication data symbols to $U$ users and emits the radar probing waveform to detect targets of interest simultaneously.
The DFRC transmitter is constructed in an HBF structure equipped with ${N_t}$ RF chains and ${M_t}$ transmit antennas. 
Besides, all the downlink users adopt the fully digital structure with $M_r$ receive antennas. 
It is assumed that both transmitter and receivers utilize the uniform linear arrays (ULAs), where the element spacing is set as $d$.

We also consider different task scenarios:
1) radar scanning and detection: the DFRC base station (BS) scans the airspace and detects any potential radar target at first.
2) target tracking: after a target is detected, the DFRC BS tracks this target in the presence of interference.

\vspace{-1.2em}
\subsection{Transmit Model}
\vspace{-0.3em}

Let $\mathbf{s}_k = [s_{k,1} , s_{k,2} , \cdots , s_{k,U}]^T \in \mathbb{C}^{U}$ be the symbols to be transmitted via the $k$-th subcarrier, $\mathbb{E} \{ {{\mathbf{s}}_k}{\mathbf{s}}_k^H \} = {\mathbf{I}}_U , k=1,\cdots, K$, where $K$ is total number of subcarriers.
The transmitted user data symbols $\mathbf{s}_k$ are firstly processed by a digital beamformer ${{\mathbf{F}}_k}  = [ {{\mathbf{f}}_{k,1}}, \cdots ,{{\mathbf{f}}_{k,U}} ]\in {\mathbb{C}^{{N_t} \times U}}$ in the frequency domain. 
After conversion to the time domain by the $KN_t$-point inverse fast Fourier transform (IFFT) and adding the cyclic prefix (CP), the signals are up-converted to the RF domain by $N_t$ RF chains.
Following this, the analog signals are beamformed by an analog beamformer $\mathbf{F}_{\mathrm{RF}} \in \mathbb{C}^{M_t \times N_t}$, which is implemented by phase shifters (PSs) to reduce the hardware cost.
Therefore, each element in $\mathbf{F}_{\mathrm{RF}}$ has a constant magnitude, i.e., $| \mathbf{F}_{\mathrm{RF}} (i,j) | = 1, \forall i,j$.
Besides, the analogy beamformer is designed for the whole bandwidth so it is the same for all the subcarriers of this system. 
% Therefore, to achieve better performance, we adopt the fully-connected structure. 
Finally, the signals are omitted by ${M_t}$ antennas.
According to above description, the transmitted time-domain signal ${\mathbf{x}}(t) \in {\mathbb{C}^{{M_t}}}$ at the time instant $t \in [0 , \Delta t]$ is given by
\begin{equation}
    {\mathbf{x}}(t) = {\mathbf{F}}_{\text{RF}} \sum\limits_{k = 1}^K  {\mathbf{F}}_k {\mathbf{s}}_k e^{\jmath 2\pi {f_k}t} ,
    \label{1}
\end{equation}
where ${f_k} = {f_c} + (k - \frac{K}{2})\Delta f$ denotes the frequency of the $k$-th subcarrier, $f_c$ denotes the center frequency, $\Delta f = \frac{B}{K} = \frac{1}{\Delta t}$ denotes subcarrier spacing, $B$ denotes the system bandwidth.

\vspace{-1em}
\subsection{Communication Model and Performance Metric}
\vspace{-0.3em}

After the transmit signal \eqref{1} propagates through the mmWave channel, the received signal ${\mathbf{r}}_{k,u}(t) \in \mathbb{C}^{{M_r}}$ of the user $u$ at   the subcarrier $k$ in time domain can be expressed as
\begin{equation}
    \begin{aligned}
        {\mathbf{r}}_{k,u}(t) & =  {\mathbf{H}}_{k,u} \mathbf{x}(t) + {{\mathbf{n}}_{k,u}}(t) \\
        & = {\mathbf{H}}_{k,u} {\mathbf{F}}_{\mathrm{RF}} {{\mathbf{F}}_k} {{\mathbf{s}}_k} {e^{\jmath2\pi {f_k}t}} + {{\mathbf{n}}_{k,u}}(t) , 
    \end{aligned}
    \label{2}
\end{equation}
where ${{\mathbf{n}}_{k,u}}(t) \in {\mathbb{C}^{{M_r}}}$ denotes additive white Gaussian noise (AWGN) of the $u$-th user at the $k$-th subcarrier, and we assume
 ${{\mathbf{n}}_{k,u}}(t) \sim \mathcal{C}\mathcal{N}\left( {0,\Delta t\sigma _n^2{{\bf{I}}_{{M_r}}}} \right)$. 
 $\mathbf{H}_{k,u} \in \mathbb{C}^{{M_r} \times {M_t}}$ denotes the wideband mmWave channel model, which can be modeled as
 \begin{equation}
     {{\bf{H}}_{k,u}} = \nu \sum\limits_{i = 0}^{L - 1} {\sum\limits_{j = 1}^{{N_{ray}}} {{\alpha _{i,j}}{{\bf{a}}_r}({\phi _{k,u,i,j}}){\bf{a}}_t^H({\theta _{k,u,i,j}}){e^{\frac{{ - \jmath2\pi i{f_k}}}{K}}}}  },
     \label{3}
 \end{equation}
where the number of clusters and rays in each cluster are denoted by $L$ and $N_{ray}$, respectively. $\nu  = \sqrt{ {{{M_t}{M_r}}} / (L{N_{ray}}) }$ is the normalization factor. ${\alpha _{i,j}}$ denotes the complex channel gain of the $j$-th ray in the $i$-th cluster.
Besides, ${{{\bf{a}}_r}({\phi _{k,u,i,j}})}$ and ${\bf{a}}_t({\theta _{k,u,i,j}})$ denote the elevation AoD/AoA of ${\phi _{k,u,i,j}}$ and ${\theta _{k,u,i,j}}$.
In this paper, we assume that the DFRC transmitter has perfect channel state information (CSI) for simplicity.

The received signal at the $u$-th user is firstly down-converted by ${N_t}$ RF chains.
Following that, every CP is removed and the signals are transformed to the frequency domain using $KN_t$-point fast Fourier transforms (FFTs).
This frequency-domain signal is then processed by a digital combiner ${\mathbf{w}_{k,u}} \in {\mathbb{C}^{{N_r}}}$ of the $u$-th user at the $k$-th subcarrier. 
Finally, the estimated symbol of the $u$-th user at the $k$-th subcarrier is given by
\begin{equation}
    \begin{aligned}
        {{\hat s}_{k,u}} =  &  {\mathbf{w}}_{k,u}^H{{\mathbf{H}}_{k,u}}{{\mathbf{F}}_{\mathrm{RF}}}{{\mathbf{f}}_{k,u}}{s_{k,u}} \\
        & + \textstyle\sum\limits_{i = 1 , i \ne u}^U \!\!\!{{\mathbf{w}}_{k,u}^H{{\mathbf{H}}_{k,u}}{{\mathbf{F}}_{\mathrm{RF}}}{{\mathbf{f}}_{k,i}}{s_{k,i}}}  + {\mathbf{w}}_{k,u}^H{{\mathbf{v}}_{k,u}} ,
    \end{aligned}
    \label{4}
\end{equation}
where ${{\mathbf{v}}_{k,u}} = \frac{1}{{\Delta t}}\int_0^{\Delta t} {{{\mathbf{n}}_{k,u}}(t){e^{ - \jmath2\pi {f_k}t}}dt}$ denotes the AWGN in frequency domain, and ${{\bf{v}}_{k,u}} \in {{\mathbb{C}}^{{M_r}}} \sim {\cal C}{\cal N}( {0,{\rm{ }}\sigma _n^2{{\bf{I}}_{{M_r}}}} )$.
Note that in equation \eqref{4}, the first term is the desired signal, the second term is the multi-user interference (MUI) and the last term is the noise part. 
According to \eqref{4}, the SINR of $u$-th user at $k$-th subcarrier is
\begin{equation}
    \begin{aligned}
        & \text{SINR}{_{k,u}}(\mathbf{F}_k , \mathbf{F}_{\mathrm{RF}} , \mathbf{w}_{k,u}) \\
        &\qquad = \frac{{| {{\mathbf{w}}_{k,u}^H{{\mathbf{H}}_{k,u}}{{\mathbf{b}}_{k,u}}} |}^2} { \textstyle\sum\limits_{i = 1 , i \ne u}^U {{{| {{\mathbf{w}}_{k,u}^H{{\mathbf{H}}_{k,u}}{{\mathbf{b}}_{k,i}}} |}^2}}  + \sigma _n^2\rm{Tr} ({\mathbf{w}}_{k,u}^H{{\mathbf{w}}_{k,u}}  )} ,
    \end{aligned}
\end{equation}
where ${{\mathbf{b}}_{k,u}} = {{\mathbf{F}}_{\mathrm{RF}}}{{\mathbf{f}}_{k,u}} \in {\mathbb{C}^{{M_t}}} , \forall k,u$.
Then, the achievable data rate of the $u$-th user at the $k$-th subcarrier is given by
\begin{equation}\label{6}
\!{R_{k,u}} (\mathbf{F}_k,\! \mathbf{F}_{\mathrm{RF}},\! \mathbf{w}_{k,u})\! = \!{\log _2}(1\! +\! \text{SINR}{_{k,u}}(\mathbf{F}_k ,\! \mathbf{F}_{\mathrm{RF}} ,\! \mathbf{w}_{k,u})) .
\end{equation}
Considering the user communication requirement, we adopt the data rate of each user on each subcarrier as a communication QoS metric.

\vspace{-0.7em}
\subsection{Radar Model and Performance Metric}
\vspace{-0.3em}

For the radar function, we focus on the radar transmit beampattern design.
Since multiple tasks are considered in the OFDM-DFRC system, it is necessary to introduce different radar beampattern metrics for different task requirements.
The first task is radar scanning and detection, which needs a wide mainlobe to scan the wide area and detect potential radar targets in this area, so it is essential to guarantee a relatively low sidelobe level and maintain the high level of the whole mainlobe region.
The second task is target tracking, which needs to concentrate energy in the target direction for tracking and reduce sidelobe clutters in the presence of interference, so it is critical to guarantee a relatively high mainlobe level and suppress the level of the whole sidelobe region.

To meet the different task requirements, we introduce two novel radar beampattern metrics to characterize the radar beampattern better in the space-frequency dimension, which is able to control the levels of both mainlobe and sidelobe more flexibly.
To be more specific, the transmit spectrum ${\cal S}_k(\mathbf{F}_{k} , \mathbf{F}_{\mathrm{RF}} , {f_k},\theta )$ towards angle $\theta $ is defined as
\begin{equation}
    \begin{aligned}
            {\cal S}_k(\mathbf{F}_{k} , \mathbf{F}_{\mathrm{RF}} , {f_k},\theta ) = & \mathbb{E}\{ {{{| {{{\mathbf{a}}^H}({f_{k}},\theta )\Delta t{{\mathbf{F}}_{\mathrm{RF}}}{{\mathbf{F}}_k}{{\mathbf{s}}_k}} |}^2}} \}  \\
            =& {(\Delta t)^2}\left\| {{\mathbf{F}}_{k}^H{\mathbf{F}}_{\mathrm{RF}}^H{\mathbf{a}}({f_{k}},\theta )} \right\|_F^2 ,
    \end{aligned}
\end{equation}
where ${\mathbf{a}}(f,\theta ) \in {\mathbb{C}^{{M_t}}}$ is the space-frequency steering vector and it can be expressed as
\begin{equation}
    {\mathbf{a}}(f,\theta ) = {\left[ {{e^{\jmath2\pi f{\tau _1}}},{e^{\jmath2\pi f{\tau _2}}}, \ldots ,{e^{\jmath2\pi f{\tau _{{M_t}}}}}} \right]^T} ,
\end{equation}
with ${\tau _m} = {{(m - 1)d\sin \theta }} / {c}$, and $c$ is the speed of electromagnetic wave.

Then, we divide the spatial angle regions into three discrete sets, i.e., mainlobe $\bm{\Theta}  = \{ {\theta _m}\} _{m = 1}^M$, sidelobe $\bm{\Omega}  = \{ {\vartheta _s}\} _{s = 1}^S$ and transition band $\bm \Phi$, where $M$ and $S$ represent the number of discrete points in the mainlobe and sidelobe angle grids, respectively.

For the scanning and detection task, the level in the mainlobe region should be maintained high and the sidelobe should be guaranteed low, so we introduce the average integrated sidelobe to minimum mainlobe ratio (AISMMR) to characterize the beampattern, which is defined as
\begin{equation}
    \mathrm{AISMMR} (\mathbf{F}_{k} , \mathbf{F}_{\mathrm{RF}}) = \frac{1}{K}\sum\limits_{k = 1}^K {\frac{{\sum\limits_{s = 1}^S {\left\| {{\bf{F}}_k^H{\bf{F}}_{{\rm{RF}}}^H{\bf{a}}({f_k},{\vartheta _s})} \right\|_F^2} }}{{\mathop {\min }\limits_{{\theta _m} \in {\bf{\Theta }}} \left\| {{\bf{F}}_k^H{\bf{F}}_{{\rm{RF}}}^H{\bf{a}}({f_k},{\theta _m})} \right\|_F^2}}} .\nonumber
\end{equation}

For the target tracking task, the level in the sidelobe should be suppressed to reduce clutters and the mainlobe level should be guaranteed high, so we introduce the average peak sidelobe to integrated mainlobe ratio (APSIMR) to characterize the beampattern, which is defined as
\begin{equation}
    \mathrm{APSIMR} (\mathbf{F}_{k} , \mathbf{F}_{\mathrm{RF}}) = \frac{1}{K}\sum\limits_{k = 1}^K {\frac{{\mathop {\max }\limits_{{\vartheta _s} \in {\bf{\Omega }}}\left\| {{\bf{F}}_k^H{\bf{F}}_{{\rm{RF}}}^H{\bf{a}}({f_k},{\vartheta _s})} \right\|_F^2}}{{\sum\limits_{m = 1}^M {\left\| {{\bf{F}}_k^H{\bf{F}}_{{\rm{RF}}}^H{\bf{a}}({f_k},{\theta _m})} \right\|_F^2} }}}.\nonumber
\end{equation}
Obviously, for each task, the smaller the APSIMR or AISMMR is, the relatively higher level the mainlobe can reach and the relatively lower level the sidelobe level can reach.

\vspace{-1.2em}
\subsection{Problem Statement}
\vspace{-0.3em}

In this paper, we aim to realize multiple radar tasks under the communication QoS constraint.
Specifically, a meaningful criterion of jointly optimizing the hybrid digital/analog beamformer $(\{{\mathbf{F}}_k\}, \mathbf{F}_{\mathrm{RF}})$ and downlink user combiners $(\{\mathbf{w}_{k,u}\})$ is to minimize the APSIMR and AISMMR respectively in different tasks while guaranteeing data rate and other hardware constraints. 
Mathematically, these two problems can be formulated into a task-oriented form as
\begin{subequations}
    \begin{align}
        \mathop {\min }\limits_{{\{{\mathbf{F}}_k}\} , {{\mathbf{F}}_{\mathrm{RF}}} , \{{\mathbf{w}}_{k,u}\}} & \;{g_i}(\{ {{\bf{F}}_k}\} ,{{\bf{F}}_{{\rm{RF}}}} )  \label{9a}\\
        \text{s.t.} \qquad &\left\| {{{\mathbf{F}}_{\mathrm{RF}}}{{\mathbf{F}}_k}} \right\|_F^2 = {P_k},\forall k\label{9b}\\
        &\left| {{{\mathbf{F}}_{\mathrm{RF}}}(i,j)} \right| = 1,\forall i,j\label{9c}\\
        &{R_{k,u}}({{\mathbf{F}}_k},{{\mathbf{F}}_{\mathrm{RF}}},{\mathbf{w}}_{k,u}) \geqslant {\chi _{k,u}},\forall k,u\label{9d} ,
    \end{align}
    \label{9}%
\end{subequations}
where the objective function is determined by the task index $i = 1,2$, which is defined as
\begin{equation} 
{g_i}(\{ {{\bf{F}}_k}\} ,{{\bf{F}}_{{\rm{RF}}}})= \left\{ \begin{array}{ll}
{\rm{AISMMR}}(\{ {{\bf{F}}_k}\},{{\bf{F}}_{{\rm{RF}}}}), & i = 1\\
 {\rm{APSIMR}}(\{ {{\bf{F}}_k}\} ,{{\bf{F}}_{{\rm{RF}}}}),& i = 2.
\end{array} \right.\nonumber
\end{equation}
The constraint \eqref{9b} denotes the transmit power of each subcarrier, 
the constraint \eqref{9c} denotes the constant modulus constraint of the analog beamformer, 
and the constraint \eqref{9d} denotes the data rate requirement of the $u$-th user on the $k$-th subcarrier with the threshold ${\chi _{k,u}}$.

The task-oriented problem \eqref{9} is a high-dimensional and non-convex optimization problem, which is hard to settle by existing methods.
Therefore, in the following, we propose a unified framework to solve it.

% \vspace{-0.5em}
\section{A Unified Optimization Approach to HBF Design for the Multi-Task DFRC System}
\vspace{-0.2em}

In this section, we first reformulate the original problem into a simple form and propose a more efficient unified optimization algorithm to jointly design the hybrid analog/digital beamformer for multiple tasks.

\vspace{-1.2em}
\subsection{Problem Reformulation}
\vspace{-0.3em}

In this subsection, to facilitate the algorithm design, we reformulate the complicated and non-smooth objective function, simplify the QoS constraints by adopting the rate-WMMSE relationship, and present the overview of the consensus alternating direction method of multipliers (CADMM) framework \cite{boyd2011distributed} to settle the problem \eqref{10} as following:

\subsubsection{Reformulation of Objective Function}
For simplification, we first introduce two sets of auxiliary variables $\{ \eta_k \}$ and $\{ \varepsilon_k \}$. 
Then, to separate $\{ \eta_k \}$ and $\{ \varepsilon_k \}$ 
in the fractional operation of the unified-framework objective function, we adopt the fractional-log smoothing approach \cite{8387476} and rewrite the problem \eqref{9} as
\begin{subequations}\label{10}
    \begin{align}
        &\mathop {\min }\limits_{\scriptstyle\{ {{\bf{w}}_{k,u}}\} ,\{ {{\bf{F}}_k}\} ,\hfill\atop
        \scriptstyle{{\bf{F}}_{{\rm{RF}}}},\{ {\eta _k}\} ,\{ {\varepsilon _k}\} \hfill} \;{{\dot g}_i}(\{ {{\bf{F}}_k}\} ,{{\bf{F}}_{{\rm{RF}}}},\{ {\eta _k}\},\{ {\varepsilon _k}\})\\
        &\qquad\; \rm{s.t.}\qquad\; \eqref{9b}-\eqref{9d}\\
        &\qquad\qquad\qquad {\rm{C}}_i(\{ {{\bf{F}}_k}\} ,{{\bf{F}}_{{\rm{RF}}}},\{ {\eta _k}\} ,\{ {\varepsilon _k}\} ),
    \end{align}
\end{subequations}
where the objective function ${{\dot g}_i}(\{ {{\bf{F}}_k}\},{{\bf{F}}_{{\rm{RF}}}}) =$
\begin{equation}
\left\{ \begin{array}{ll}
\frac{1}{K}\sum\limits_{k = 1}^K {(\log (\sum\limits_{s = 1}^S  {\bf{F}}_k^H{\bf{F}}_{{\rm{RF}}}^H{\bf{a}}({f_k},{\vartheta _s})_F^2) - \log ({\varepsilon _k}))} ,& i = 1\\
\frac{1}{K}\sum\limits_{k = 1}^K {(\log ({\eta _k}) - \log (\sum\limits_{m = 1}^M  {\bf{F}}_k^H{\bf{F}}_{{\rm{RF}}}^H{\bf{a}}({f_k},{\theta _m})_F^2))}, & i = 2,
\end{array} \right.\nonumber
\end{equation}
and the inequality constraint ${\rm{C}}_i(\{ {{\bf{F}}_k}\} ,{{\bf{F}}_{{\rm{RF}}}},\{ {\eta _k}\} ,\{ {\varepsilon _k}\} ) : $
\begin{equation}
\left\{ \begin{array}{ll}
\mathop {\min }\nolimits_{{\theta _m} \in {\bf{\Theta }}} {\|{\bf{F}}_k^H{\bf{F}}_{{\rm{RF}}}^H{\bf{a}}({f_k},{\vartheta _s})_F^2\|} \ge {\varepsilon _k},\forall k,& i = 1\\
\mathop {\max }\nolimits_{{\vartheta _s} \in {\bf{\Omega }}} {\|{\bf{F}}_k^H{\bf{F}}_{{\rm{RF}}}^H{\bf{a}}({f_k},{\vartheta _s})_F^2\|} \le {\eta _k},\forall k,& i = 2,
\end{array} \right.\nonumber
\end{equation}
are both determined by the tasks.
The problem \eqref{10} is more tractable than the problem \eqref{9} since the complicated max-min fractional objective function is removed.

\subsubsection{Rate-WMMSE Transformation}
To tackle the difficulty arising from the QoS constraints \eqref{9d}, we consider adopting the rate-WMMSE relationship, to which we give the following useful definition.
\begin{defi}\label{defi_1}
According to \eqref{4}, the mean square error (MSE) matrix function is expressed as
\begin{equation}
    \begin{aligned}
        & {e_{k,u}} (\mathbf{F}_k , \mathbf{F}_{\mathrm{RF}} , \mathbf{w}_{u,k} ) = \mathbb{E}\{ {{{\left| {{{\hat s}_{k,u}} - {s_{k,u}}} \right|}^2}} \}  \\
        &= \!\! {| {{\mathbf{w}}_{k,u}^H{{\mathbf{H}}_{k,u}}{{\mathbf{b}}_{k,u}} \!- \!1} |^2} \!\! 
        +\!\!\!\! \!\!\!\textstyle\sum\limits_{i = 1 , i \ne u}^U\!\!\!\! \!\! {{{| {{\mathbf{w}}_{k,u}^H{{\mathbf{H}}_{k,u}}{{\mathbf{b}}_{k,i}}} |}^2}} \!\!\! + \! \sigma _n^2{\mathbf{w}}_{k,u}^H{{\mathbf{w}}_{k,u}} .\nonumber
    \end{aligned}
    \label{14}
\end{equation}
\end{defi}
Then, based on definition \ref{defi_1}, we can build the relationship between rate and weighted MSE minimization using the following proposition.
\begin{prop}\label{prop_1}
    By introducing auxiliary weighting coefficients $\{ \omega _{k,u} \}$, the rate ${R_{k,u}}(\mathbf{F}_k , \mathbf{F}_{\mathrm{RF}} , \mathbf{w}_{u,k} )$  can be equivalently expressed as
    \begin{equation}
        {R_{k,u}}(\mathbf{F}_k , \mathbf{F}_{\mathrm{RF}} , \mathbf{w}_{u,k} ) = {{{\log }_2}({\omega _{k,u}}) - {\omega _{k,u}}{e_{k,u}} + 1},
        \label{15} 
    \end{equation}
    where auxiliary variable $\omega _{k,u}$ can be updated by $\omega _{k,u} = 1/e_{k,u}$, and $\mathbf{w}_{k,u}$ can be updated by 
    \begin{equation}
        {\mathbf{w}}_{k,u}^{\mathrm{MMSE}} = {\left( {{{\mathbf{H}}_{k,u}}{{\mathbf{F}}_{\mathrm{RF}}}{{\mathbf{F}}_k}{\mathbf{F}}_k^H{\mathbf{F}}_{\mathrm{RF}}^H{\mathbf{H}}_{k,u}^H + \sigma _n^2{{\mathbf{I}}_{{M_r}}}} \right)^{ - 1}}{{\mathbf{H}}_{k,u}}{{\mathbf{b}}_{k,u}}.
    \end{equation}
\end{prop}
\begin{IEEEproof}
    Please refer to \cite{WMMSE2015Jia}.
\end{IEEEproof}

Then, according to proposition \ref{prop_1}, the user date rate equation \eqref{6} can be equivalently converted into
\begin{equation}
    {R_{k,u}}( {{\mathbf{F}}_k} , {{\mathbf{F}}_{\mathrm{RF}}} , {\mathbf{w}}_{k,u} )  =  {{{\log }_2}({\omega _{k,u}}) -  {\omega _{k,u}}{e_{k,u}} + 1}. \label{16}
\end{equation}

\subsubsection{Application of CADMM Framework}

Based on the above reformulation, the problem \eqref{10} can be further rewritten as
\begin{subequations}\label{A1}
    \begin{align}
        &\mathop {\min }\limits_{\scriptstyle\{ {{\bf{F}}_k}\} ,{{\bf{F}}_{{\rm{RF}}}},\hfill\atop
        \scriptstyle\{ {\eta _k}\} ,\{ {\varepsilon _k}\} \hfill} \;{{\dot g}_i}(\{ {{\bf{F}}_k}\} ,{{\bf{F}}_{{\rm{RF}}}},\{ {\eta _k}\},\{ {\varepsilon _k}\}) \label{A1_1}\\
        &\quad\;\;\rm{s.t.}\quad\;\; \left\| {{{\bf{F}}_{{\rm{RF}}}}{{\bf{F}}_k}} \right\|_F^2 = {P_k},\forall k \label{A1_2}\\
        &\qquad\qquad\;\; \left| {{{\bf{F}}_{{\rm{RF}}}}(i,j)} \right| = 1,\forall i,j \label{A1_3}\\
        &\qquad\qquad\;\;{{{\log }_2}({\omega _{k,u}})\! - \! {\omega _{k,u}}{e_{k,u}}\! +\! 1} \ge {\chi _{k,u}},\forall k,u \label{A1_4}\\
        &\qquad\qquad\;\; {\rm{C}}_i(\{ {{\bf{F}}_k}\} ,{{\bf{F}}_{{\rm{RF}}}},\{ {\eta _k}\} ,\{ {\varepsilon _k}\} ), \label{A1_5}
    \end{align}
\end{subequations}
Although the problem \eqref{A1} is more tractable than the original problem \eqref{9}, it is still hard to solve by the CADMM method since there are coupling constraints for different variables.
To decouple the hybrid digital/analog beamformers and solve the troublesome constraints \eqref{A1_2}-\eqref{A1_5} from others, we introduce several auxiliary variables and rewrite the problem \eqref{A1} equivalently as follows.
\begin{subequations}
    \begin{align}
        &\mathop {\min }\limits_{\scriptstyle\;\{ {{\bf{Y}}_k}\} ,\{ {{\bf{F}}_k}\} ,{{\bf{F}}_{{\rm{RF}}}},\hfill\atop
        {\scriptstyle\{ {\eta _k}\} ,\{ {\varepsilon _k}\} ,\{ G_{k,u}^{k,i}\} ,\hfill\atop\quad\{ {{\bf{h}}_{k,s}}\} ,\{ {{\bf{g}}_{k,m}}\} \hfill}} \;{{\ddot g}_i}(\{ {{\bf{h}}_{k,s}}\} ,\{ {{\bf{g}}_{k,m}}\} ,\{ {\eta _k}\} ,\{ {\varepsilon _k}\} ) \label{A2_a}\\
        &\qquad\quad\rm{s.t.}\;\;{\left\| {{{\bf{Y}}_k}} \right\|_F^2 = {P_k},\forall k}\label{A2_b}\\
        &\qquad\qquad\quad{\left| {{{\bf{F}}_{{\rm{RF}}}}(i,j)} \right| = 1,\forall i,j}\label{A2_c}\\
        &\qquad\qquad\quad|G_{k,u}^{k,u} - 1{|^2} +\!\!\!\! \textstyle\sum\limits_{i = 1.i \ne u}^U\!\!\!\! {|G_{k,u}^{k,i}{|^2}}  \le {\xi _{k,u}},\forall k,u\label{A2_d}\\
        &\qquad\qquad\quad{{{\rm{\dot C}}}_i}(\{ {{\bf{h}}_{k,s}}\} ,\{ {{\bf{g}}_{k,m}}\} ,\{ {\eta _k}\} ,\{ {\varepsilon _k}\} )\label{A2_e}\\
        &\qquad\qquad\quad{{\bf{Y}}_k} = {{\bf{F}}_{{\rm{RF}}}}{{\bf{F}}_k},\forall k \label{A2_f}\\
        &\qquad\qquad\quad G_{k,u}^{k,i} = {\bf{w}}_{k,u}^H{{\bf{H}}_{k,u}}{{\bf{y}}_{k,i}},\forall k,i,u \label{A2_g}\\
        &\qquad\qquad\quad{{\bf{h}}_{k,s}} = {\bf{Y}}_k^H{\bf{a}}({f_k},{\vartheta _s}),\forall k,s \label{A2_h}\\
        &\qquad\qquad\quad{{\bf{g}}_{k,m}} = {\bf{Y}}_k^H{\bf{a}}({f_k},{\theta _m}),\forall k,m ,\label{A2_i}
    \end{align}\label{A2}%
\end{subequations}
where  ${\xi _{k,u}} = ({\log _2}({\omega _{k,u}}) - {\omega _{k,u}}\sigma _n^2{\bf{w}}_{k,u}^H{{\bf{w}}_{k,u}} + 1 - {\chi _{k,u}})/{\omega _{k,u}}$.
The objective function ${{\ddot g}_i}(\{ {{\bf{h}}_{k,s}}\} ,\{ {{\bf{g}}_{k,m}}\} ,\{ {\eta _k}\} ,\{ {\varepsilon _k}\} ) = $
\begin{equation}
    \left\{ \begin{array}{ll}
    \frac{1}{K}\sum\limits_{k = 1}^K {(\log (\sum\limits_{s = 1}^S  \left\| {{{\bf{h}}_{k,s}}} \right\|_F^2) - \log ({\varepsilon _k}))},&i = 1 \\
    \frac{1}{K}\sum\limits_{k = 1}^K {(\log ({\eta _k}) - \log (\sum\limits_{m = 1}^M  \left\| {{{\bf{g}}_{k,m}}} \right\|_F^2))},&i = 2
\end{array} \right.\nonumber
\end{equation}
and the inequality constraint ${{{\rm{\dot C}}}_i}(\{ {{\bf{h}}_{k,s}}\} ,\{ {{\bf{g}}_{k,m}}\} ,\{ {\eta _k}\} ,\{ {\varepsilon _k}\} ):$
\begin{equation}
    \left\{ \begin{array}{ll}
    \mathop {\min }\nolimits_{{\theta _m} \in {\bf{\Theta }}} \left\| {{{\bf{g}}_{k,m}}} \right\|_F^2 \ge {\varepsilon _k},\forall k,&i = 1\\
    \mathop {\max }\nolimits_{{\vartheta _s} \in {\bf{\Omega }}} \left\| {{{\bf{h}}_{k,s}}} \right\|_F^2 \le {\eta _k},\forall k,&i = 2
\end{array} \right.\nonumber
\end{equation}
are determined by the tasks. 
\begin{figure*}
 	\centering
        \begin{equation}
            \begin{aligned}
                {{\cal L}_{i,k}}&({{\bf{Y}}_k},{{\bf{F}}_k},{{\bf{F}}_{{\rm{RF}}}}, {\eta _k},{\varepsilon _k},\{ G_{k,u}^{k,i}\} ,\{ {{\bf{h}}_{k,s}}\} ,\{ {{\bf{g}}_{k,m}}\} ) \\
                =& {{\ddot g}_i}(\{ {{\bf{h}}_{k,s}}\} ,\{ {{\bf{g}}_{k,m}}\} ,\{ {\eta _k}\} ,\{ {\varepsilon _k}\} ) + {\frac{{{\rho _{1,k}}}}{2}}{\textstyle\sum\limits_{u = 1}^U {\left\| {{{\bf{y}}_{k,u}} - {{\bf{b}}_{k,u}} + {{\bm{\varsigma }}_{k,u}}} \right\|_F^2} }  + {\frac{{{\rho _{2,k}}}}{2}\textstyle\sum\limits_{u = 1}^U{\textstyle\sum\limits_{i = 1}^U {\left\| {G_{k,u}^{k,i} - {\bf{w}}_{k,u}^H{{\bf{H}}_{k,u}}{{\bf{y}}_{k,i}} + {\bf{\lambda }}_{k,u}^{k,i}} \right\|_F^2} }} \\
                &+ {\frac{{{\rho _{3,k}}}}{2}\textstyle\sum\limits_{s = 1}^S {\left\| {{{\bf{h}}_{k,s}} - {\bf{Y}}_k^H{\bf{a}}({f_k},{\vartheta _s}) + {{\bm{\beta }}_{k,s}}} \right\|_F^2}} +\frac{{{\rho _{4,k}}}}{2}\textstyle\sum\limits_{m = 1}^M {\left\| {{{\bf{g}}_{k,m}} - {\bf{Y}}_k^H{\bf{a}}({f_k},{\theta _m}) + {{\bm{\nu }}_{k,m}}} \right\|_F^2}
            \end{aligned}
            \label{AL}
        \end{equation}
        \hrule
    \vspace{-1.5em}
\end{figure*}
The associated augmented Lagrangian (AL) function via penalizing the equality constraints \eqref{A2_f}-\eqref{A2_i} is given by
\begin{equation}
{{\cal L}_i}\! =\! \frac{1}{K}\!\!\sum\limits_{k = 1}^K\! {{{\cal L}_{i,k}}({{\bf{Y}}_{k}},{{\bf{F}}_{k}},{{\bf{F}}_{{\rm{RF}}}},{\eta _k},{\varepsilon _k},\{ G_{k,u}^{k,i}\},\{ {{\bf{h}}_{k,s}}\},\{ {{\bf{g}}_{k,m}}\} )},\nonumber
\end{equation}
where ${{{\cal L}_{i,k}}({{\bf{Y}}_{k}},{{\bf{F}}_{k}},{{\bf{F}}_{{\rm{RF}}}},{\eta _k},{\varepsilon _k},\{ G_{k,u}^{k,i}\},\{ {{\bf{h}}_{k,s}}\},\{ {{\bf{g}}_{k,m}}\} )}$ is defined in the equation \eqref{AL}, and the value of $i$ is also determined by the tasks.
${\rho _{1,k}},~{\rho _{2,k}},~{\rho _{3,k}},~{\rho _{4,k}} > 0$ are the corresponding penalty parameters. 
$\{ {{{\bm{\varsigma }}_{k,u}}} \} $, $\{ { \lambda _{k,u}^{k,i}} \}$, $\{ {{{\bm{\beta }}_{k,s}}} \} $ and $\{ {{{\bm{\nu }}_{k,m}}} \} $ are dual variables for constraints \eqref{A2_f}-\eqref{A2_i}, respectively.
By doing so, under the CADMM framework, the resultant problem \eqref{A2} can be solved by taking iterations steps in the following subsection.

\vspace{-1.5em}
\subsection{ Solutions to Unified Task-Oriented Optimization Problem}
\vspace{-0.4em}

\subsubsection{Optimize ${{{\bf{Y}}_k}}$ } 
With given other variables, ${\bf{Y}}_{{k}}$ can be updated by solving the subproblem:
\begin{equation}
    \begin{aligned}
    &\mathop {\min }\limits_{\mathbf{Y}_k} \;\;\frac{{{\rho _{1,k}}}}{2}\left\| {{\mathbf{Y}_k} - {\mathbf{M}_k}} \right\|_F^2\\ 
    & \qquad\;\;+  \frac{{{\rho _{2,k}}}}{2}{\sum\limits_{u = 1}^U {\sum\limits_{i = 1}^U {{{\left| {G_{k,u}^{k,i} - {\mathbf{w}}_{k,u}^H{{\mathbf{H}}_{k,u}}{\mathbf{y}}_{k,i} + \lambda _{k,u}^{k,i}} \right|}^2}} } } \\
    & \qquad\;\;+  \frac{{{\rho _{3,k}}}}{2}{\sum\limits_{s = 1}^S \!{\left\| {{{\mathbf{h}}_{k,s}} - {\mathbf{Y}}_k^H{\mathbf{a}}({f_{{k}}},{\vartheta _s}) + {{\bm{\beta }}_{k,s}}} \right\|_F^2} } \\ 
    & \qquad\;\;+  \frac{{{\rho _{4,k}}}}{2}{\sum\limits_{m = 1}^M\! {\left\| {{{\mathbf{g}}_{k,m}} - {\mathbf{Y}}_k^H{\mathbf{a}}({f_{{k}}},{\theta _m}) + {{\bm{\nu }}_{k,m}}} \right\|_F^2} } \\ 
    &{\text{ s.t.}}\;\;\;\;\;\mathrm{Tr}( {{{\mathbf{Y}}_k}{\mathbf{Y}}_k^H} ) = {P_k} ,
   \end{aligned}
   \label{22}
\end{equation}
where ${{\mathbf{M}}_k} = [{{\mathbf{b}}_{k,1}} - {{\bm{\varsigma }}_{k,1}}, \ldots ,{{\bm{b}}_{k,U}} - {{\bm{\varsigma }}_{k,U}}]$.
The problem \eqref{22} is a quadratic program problem with equality constraint, which can be solved by analyzing the Karush-Kuhn-Tucker (KKT) conditions.
Specifically, the closed-form solution to $\mathbf{Y}_k$ can be derived as
\begin{equation}
    {{\bf{Y}}_k}({\mu _k}) = ({{\bf{\Xi }}_k + \mu_k \mathbf{I}_{M_t} )^{ - 1}} \times {\bf{\Psi }}_k,
    \label{21}
\end{equation}
where ${\mu _k}$ is a multiplier for the constraint $\mathrm{Tr}( {{{\mathbf{Y}}_k}{\mathbf{Y}}_k^H} ) = {P_k}$, the $(i,u)$-th entry of ${\bf{G}}_k^\lambda  \in {\mathbb{C}^{U \times U}}$ is $G_{k,u}^{k,i} + \lambda _{k,u}^{k,i}$, and
\begin{equation}
\begin{aligned}
    &{\bf{\Xi }}_k = \textstyle\frac{{{\rho _{1,k}}}}{2}{{\bf{I}}_{{M_t}}} + \frac{{{\rho _{2,k}}}}{2}{{\bf{B}}_k}{\bf{B}}_k^H + \frac{{{\rho _{3,k}}}}{2}{{\bf{A}}_S} + \frac{{{\rho _{4,k}}}}{2}{{\bf{A}}_M},\\ 
    &{\bf{\Psi }}_k =\textstyle \frac{{{\rho _{1,k}}}}{2}{{\bf{M}}_k} + \frac{{{\rho _{2,k}}}}{2}{{\bf{B}}_k}{\bf{G}}_k^\lambda  + \frac{{{\rho _{3,k}}}}{2}{{\bf{\Omega }}_S} + \frac{{{\rho _{4,k}}}}{2}{{\bf{\Omega }}_M},\\
    &{{\bf{B}}_k} = [ {\bf{H}}_{k,1}^H{{\bf{w}}_{k,1}}, \ldots ,{\bf{H}}_{k,U}^H{{\bf{w}}_{k,U}} ],\\ 
    &{{\bf{A}}_S} \!=\!\! \textstyle\sum\limits_{s = 1}^S \!{{\bf{a}}({f_{{k}}},\!{\vartheta _s}){{\bf{a}}^H}({f_{{k}}},\!{\vartheta _s})},
    {{\bf{A}}_M} \!=\!\! \textstyle\sum\limits_{m = 1}^M \!\!{{\bf{a}}({f_{{k}}},\!{\theta _m}){{\bf{a}}^H}({f_{{k}}},\!{\theta _m})},\\
    &{{\bf{\Omega }}_S} = \textstyle\sum\limits_{s = 1}^S {{\bf{a}}({f_{{k}}},{\vartheta _s}){\bf{q}}_{k,s}^H},
    {{\bf{\Omega }}_M} = \textstyle\sum\limits_{m = 1}^M {{\bf{a}}({f_{{k}}},{\theta _m}){\bf{m}}_{k,m}^H},\\
    &{{\mathbf{q}}_{k,s}} = {{\mathbf{h}}_{k,s}} + {{\bm{\beta }}_{k,s}} \in {\mathbb{C}^{U}},
    {{\mathbf{m}}_{k,m}} = {{\mathbf{g}}_{k,m}} + {{\bm{\nu }}_{k,m}} \in {\mathbb{C}^{U}}.    
\end{aligned}\nonumber
\end{equation}

To obtain the optimal solution to $\mathbf{Y}_{k}$, we need to find the optimal solution to multiplier ${\mu _k}$.
To this end, we define the eigen-decomposition of ${\bf{\Xi }}_k$ as ${\bf{\Xi }}_k = {\mathbf{D}}_k {\bm \Sigma}_k {{\mathbf{D}}_k^H}$.
Then, equation \eqref{21} can be replaced by
\begin{equation}
    {{\mathbf{Y}}_k}({\mu _k}) = {\mathbf{D}}_k {\left( {{\bm{\Sigma }}_k + {\mu _k}{{\mathbf{I}}_{{M_t}}}} \right)^{ - 1}}{{\mathbf{D}}_k^H}{\mathbf{\Psi }}_k.
    \label{W_23}
\end{equation}
By substituting \eqref{W_23} into the constraint of power $\mathrm{Tr}( {{{\mathbf{Y}}_k}{\mathbf{Y}}_k^H} ) = {P_k}$, we get
\begin{equation}
    \mathrm{Tr}\left( {{{\mathbf{Y}}_k}{\mathbf{Y}}_k^H} \right) = \sum\limits_{m = 1}^{M_t} {\frac{{{{\left[ {{\mathbf{D}}_k^H}{\bm{\Psi}_k}{{\bm{\Psi}}_k^H}{\mathbf{D}_k} \right]}_{m,m}}}}{{{{( {{\mu _k} + {{\left[ {\bm{\Sigma}_k} \right]}_{m,m}}} )}^2}}}}  = {P_k}.
\end{equation}
Thus, the optimal $\mu _k^{\star}$ can be determined by bisection or Newton’s method.
Finally, substituting $\mu _k^{\star}$ into ${{\mathbf{Y}}_k}({\mu _k})$, the optimal solution can be obtained by ${\mathbf{Y}}_k^{\star} = {{\mathbf{Y}}_k}(\mu _k^{\star})$.

\subsubsection{Optimize ${\{ {G_{k,u}^{k,i}} \}}$} 

With given other variables, ${\{ {G_{k,u}^{k,i}} \}}$ can be updated by solving the subproblem:
\begin{subequations}\label{31}
\begin{align}
    &\mathop {\min }\limits_{\left\{ {G_{k,u}^{k,i}} \right\}} \sum\limits_{u = 1}^U {\sum\limits_{i = 1}^U {{{\left| {G_{k,u}^{k,i} - {\mathbf{w}}_{k,u}^H{{\mathbf{H}}_{k,u}}{\mathbf{y}}_{k,i}^{} + \lambda _{k,u}^{k,i}} \right|}^2}} } \\
  &\quad{\text{s.t. }}\;\;|G_{k,u}^{k,u} - 1{|^2} + \!\!\!\! \sum\limits_{i = 1.i \ne u}^U \!\!\!\!{|G_{k,u}^{k,i}{|^2}}  \le {\xi _{k,u}},\forall k,u.
\end{align}
\end{subequations}
The problem \eqref{31} is a quadratic optimization problem, which can be solved by analyzing the KKT conditions. Specifically, we first derive the Lagrange function as:
\begin{equation}
  \begin{aligned}
   &\!\!\widetilde {\mathcal{L}}( {\{ {G_{k,u}^{k,i}} \},\varphi } )\! =\! \textstyle\sum\limits_{u = 1}^U {\sum\limits_{i = 1}^U {{{| {G_{k,u}^{k,i} - {\mathbf{w}}_{k,u}^H{{\mathbf{H}}_{k,u}}{\mathbf{y}}_{k,i}^{} + \lambda _{k,u}^{k,i}} |}^2}} } \\
   & + \varphi ( |G_{k,u}^{k,u} - 1{|^2} + \!\!\!\! \textstyle\sum\limits_{i = 1.i \ne u}^U \!\!\!\!{|G_{k,u}^{k,i}{|^2}} - {\xi _{k,u}}),
  \end{aligned}
\end{equation}\label{32}%
where $\varphi  \geqslant 0$ is a Lagrange multiplier. Then, according to the first-order optimality condition, $G_{k,u}^{k,i}$ can be determined by
\begin{equation}
G_{k,u}^{k,i} (\varphi) = 
\left\{ \begin{array}{ll}
    \frac{{{\bf{w}}_{k,u}^H{{\bf{H}}_{k,u}}{{\bf{y}}_{k,i}} - \lambda _{k,u}^{k,i} + \varphi }}{{1 + \varphi }} ,&u = i\\
    \frac{{{\bf{w}}_{k,u}^H{{\bf{H}}_{k,u}}{{\bf{y}}_{k,i}} - \lambda _{k,u}^{k,i}}}{{1 + \varphi }},& u \ne i
\end{array} \right.
    \label{33}
\end{equation}
Based on the value of $\varphi$, we have the following two cases:\\
$\bullet$ Case 1: $\varphi  = 0$

Submitting $\varphi  = 0$ into \eqref{33}, we can obtain
\begin{equation}\label{34}
        G_{k,u}^{k,i}(0) = {\mathbf{w}}_{k,u}^H{{\mathbf{H}}_{k,u}}{{\mathbf{y}}_{k,i}} - \lambda _{k,u}^{k,i}.
\end{equation}
If $|G_{k,u}^{k,u} - 1{|^2} +  \sum\nolimits_{i = 1.i \ne u}^U  {|G_{k,u}^{k,i}{|^2}}  \le {\xi _{k,u}}$ can be satisfied, the optimal solution is ${\{ {G_{k,u}^{k,i}} \}^{\star}} = \{ {G_{k,u}^{k,i}(0)} \}$.\\
$\bullet$ Case 2: $\varphi  > 0$

In this case, we have the following relationship:
\begin{equation}\label{35}
    |G_{k,u}^{k,u} - 1{|^2} + \!\!\!\!\sum\limits_{i = 1.i \ne u}^U \!\!\!\!\! {|G_{k,u}^{k,i}{|^2}}  = {\xi _{k,u}},
\end{equation}
whose optimal root ${\varphi ^{\star}}$ can be found using the Newton's method \cite{wood_1992}. 
Plugging optimal ${\varphi ^{\star}}$ into \eqref{33}, the optimal solution is ${\{ {G_{k,u}^{k,i}} \}^{\star}} = \{ {G_{k,u}^{k,i}({\varphi ^{\star}})} \}$.

\subsubsection{Optimize $\{ {{\bf{h}}_{k,s}},{\eta _k}\}$ } With given other variables, $\{ {{\bf{h}}_{k,s}},{\eta _k}\}$ can be updated by solving the subproblems.

\noindent 
(3-i) When $i=1$: the subproblem is based on minimizing the integrated sidelobe level.
\begin{equation}\label{IS_1}
\mathop {\min }\limits_{\{ {\bf{h}}_{k,s}\} }\; \log ( {\sum\limits_{s = 1}^S {\| {{{\bf{h}}_{k,s}}} \|_F^2} } ) + \frac{{{\rho _{3,k}}}}{2}\sum\limits_{s = 1}^S {\| {{{\bf{h}}_{k,s}} - \widehat {\bf{h}}_{k,s}^{}} \|_F^2},
\end{equation}
where $\widehat {\bf{h}}_{k,s}^{} = {\bf{Y}}_k^H{\bf{a}}({f_{k}},{\vartheta _s}) - {\bm{\beta }}_{k,s}^{},\forall s,k$.
Note that we do not need to optimize ${\eta _k}$ here since the objective function in this subproblem is not related to ${\eta _k}$.
Then, for simplification, \eqref{IS_1} can be written in an equivalent form:
\begin{equation}\label{IS_2}
    \mathop {\min }\limits_{{{\bf{r}}_k}} \quad \log \| {{{\bf{r}}_k}} \|_F^2 + \frac{{{\rho _{3,k}}}}{2}\| {{{\bf{r}}_k} - {{{\bf{\hat r}}}_k}} \|_F^2,
\end{equation}
where ${{\bf{r}}_k} = [{{\bf{h}}_{k,1}}; \ldots ;{{\bf{h}}_{k,S}}]$ and ${{{\bf{\hat r}}}_k} = [{{{\bf{\hat h}}}_{k,1}}; \ldots ;{{{\bf{\hat h}}}_{k,S}}]$.

Then, we approximate the objective function in \eqref{IS_2} by taking the first-order Taylor expansion of the first term in \eqref{IS_2} with respect to ${{\bf{r}}_k}$ at its last iteration point ${{{{\bf{\bar r}}}_k}}$,
which can be expressed as
\begin{equation}\label{IS_3}
    \mathop {\min }\limits_{{{\bf{r}}_k}} \; {\log \| {{{{\bf{\bar r}}}_k}} \|_F^2 + \frac{{{{{\bf{\bar r}}}_k}}}{{\| {{{{\bf{\bar r}}}_k}} \|_F^2}}( {{{\bf{r}}_k} - {{{\bf{\bar r}}}_k}} )}  + \frac{{{\rho _{3,k}}}}{2}\| {{{\bf{r}}_k} - {{{\bf{\hat r}}}_k}} \|_F^2,
\end{equation}
Since the problem \eqref{IS_3} is an unconstrained optimization problem, its optimal solution can be easily derived as ${{\bf{r}}_k^{\star}} = {{{\bf{\hat r}}}_k} - {{{{{\bf{\bar r}}}_k}}}/({{{\rho _{3,k}}\| {{{{\bf{\bar r}}}_k}} \|_F^2}})$.
Accordingly, the optimal ${\mathbf{h}}_{k,s}^{\star}$ can be obtained.

\noindent 
(3-ii) When $i=2$: the subproblem is based on minimizing the peak sidelobe level.
\begin{subequations}\label{36}
\begin{align}
&\mathop {\min }\limits_{\{ {\bf{h}}_{k,s}^{},{\eta _k}\} } \log ({\eta _k}) + \frac{{{\rho _{3,k}}}}{2}\sum\limits_{s = 1}^S {\| {{{\bf{h}}_{k,s}} - \widehat {\bf{h}}_{k,s}^{}} \|_F^2} \label{37a}\\
&\quad\;{\rm{s.t.}}\quad\| {{{\bf{h}}_{k,s}}}\|_F^2 \le {\eta _k},{\rm{ }}\forall s,{{k}}\label{37b},
\end{align}
\end{subequations}
where $\widehat {\bf{h}}_{k,s}^{} = {\bf{Y}}_k^H{\bf{a}}({f_{k}},{\vartheta _s}) - {\bm{\beta }}_{k,s}^{},\forall s,k$.

It can be observed that ${\eta}_k$ and ${{\mathbf{h}}_{k,s}}$ are coupled under the constraint $\| {{{\bf{h}}_{k,s}}} \|_F^2 \le {\eta _k}$. Therefore, if $\eta _k^{\star}$ is known, the optimal ${\mathbf{h}}_{k,s}^{\star}$ can be determined by solving
\begin{subequations}
    \begin{align}
    &\mathop {\min }\limits_{\{ {{\bf{h}}_{k,s}}\} } \sum\limits_{s = 1}^S {\| {{{\bf{h}}_{k,s}} - \widehat {\bf{h}}_{k,s}^{}} \|_F^2} \\
    &\;\;\;{\rm{s.t.}}\;\left\| {{{\bf{h}}_{k,s}}} \right\|_F^2 \le \eta _k^{\star},\forall s,
    \end{align}\label{37}
\end{subequations}
whose solution can be calculated as
\begin{equation}\label{38}
{\bf{h}}_{k,s}^{\star} = \left\{ \begin{array}{ll}
    \sqrt {\eta _k^{\star}} \frac{{\widehat {\bf{h}}_{k,s}^{}}}{{{{\| {\widehat {\bf{h}}_{k,s}^{}}\|}_F}}},&{\| {\widehat {\bf{h}}_{k,s}^{}} \|_F} > \sqrt {\eta _k^{\star}}\\
    \widehat {\bf{h}}_{k,s}, &{\rm{ otherwise}}.
\end{array} \right.
\end{equation}

Substituting \eqref{38} into \eqref{36}, we can change the initial problem to the optimization problem about a function dependent on a single variable ${\eta}_k $:
\begin{equation}
    \mathop {\min }\limits_{{\eta _k}} f_k({\eta _k}) = \log ({\eta _k}) + \frac{{{\rho _{3,k}}}}{2}\sum\limits_{s = 1}^S {{\Gamma _s}{{( {\sqrt {{\eta _k}}  - {{\| {\widehat {\bf{h}}_{k,s}^{}} \|}_F}} )}^2}},
    \label{39b}
\end{equation}
where ${\Gamma _{k,s}} = 0$ if ${\| {\widehat {\bf{h}}_{k,s}^{}} \|_F} \le \sqrt {{\eta _k}}$, otherwise, ${\Gamma _{k,s}} = 1$.

We define $[ {{{\widehat \eta }_0}, \ldots ,{{\widehat \eta }_Q}} ]$ be the ascending order sequence of $[ {{{\| {\widehat {\bf{h}}_{k,1}^{}} \|}_F}, \ldots ,{{\| {\widehat {\bf{h}}_{k,S}^{}} \|}_F}} ]$ without repetitive terms, and $Q \leqslant S$. Then, \eqref{39b} can be represented as a piece-wise function:
\begin{equation}
     {f_k}({\eta _k}) = \{ {{f_{k,q}}({\eta _k})|{{\widehat \eta }_{q - 1}} \le \sqrt {{\eta _k}}  \le {{\widehat \eta }_q},\; \forall q} \}, \label{40a}
\end{equation}
where ${f_{k,q}}({\eta _k}) = \log ({\eta _k}) + \frac{{{\rho _{3,k}}}}{2}\sum\nolimits_{n = q}^Q {{{( {\sqrt {{\eta _k}}  - {{\widehat \eta }_n}} )}^2}}  $ is defined on the interval $[ {{{\widehat \eta }_{q - 1}},{{\widehat \eta }_q}} ]$.

Since $\log ({\eta}_k )$ is non-convex and $\frac{{{\rho _{3,k}}}}{2}\sum\nolimits_{n = q}^Q {{{( {\sqrt {{\eta _k}}  - {{\widehat \eta }_n}} )}^2}}$ is convex, ${f_{k,q}}({\eta _k})$ is non-convex. To solve this non-convex problem, an auxiliary variable $\kappa  = \sqrt {{\eta _k}}  > 0$ is introduced and ${f_{k,q}}({\eta _k})$ can be transformed into:
\begin{equation}\label{41}
    {f_{k,q}}(\kappa ) = 2\log (\kappa ) + \frac{{{\rho _{3,k}}}}{2}\sum\limits_{n = q}^Q {{\kappa ^2}}  - {\rho _{3,k}}\sum\limits_{n = q}^Q {{{\widehat \eta }_n}\kappa }  + \frac{{{\rho _{3,k}}}}{2}\sum\limits_{n = q}^Q {{{\widehat \eta }_n}^2} .
    \nonumber
\end{equation}
We can easily calculate that
\begin{equation}
\left\{ \begin{array}{ll}
    {{f'}_{k,q}}(\kappa ) = 2{\kappa ^{ - 1}} + {\rho _{3,k}}\sum\limits_{n = q}^Q \kappa   - {\rho _{3,k}}\sum\limits_{n = q}^Q {{{\widehat \eta }_n}} \\
    {{f'''}_{k,q}}(\kappa ) = 4{\kappa ^{ - 3}} > 0. 
\end{array} \right.
\end{equation}
Therefore, ${f'_{k,q}}(\kappa )$ is convex on subregion $\kappa  \in [ {{{\widehat \eta }_{q - 1}},{{\widehat \eta }_q}} ]$.
To find the minimal value of ${f_{k,q}}(\kappa )$, we set ${f'_{k,q}}(\kappa ) = 0$, which can be represented as a quadratic equation with one unknown variable:
\begin{equation}\label{43}
    a{\kappa ^2} + b\kappa  + c = 0,
\end{equation}
where $a = {\rho _3}\sum\nolimits_{n = q}^Q 1  \geqslant 0$, $b =  - {\rho _3}\sum\nolimits_{n = q}^Q {{{\widehat \eta }_n}}$ and $c = 2$.

In order to analyze \eqref{41}, we should discuss the conditions for the roots of \eqref{43}. 
If we could have two real roots noted as ${v_1}$ and ${v_2}$, which can be expressed as:
\begin{equation}
    {v_1} = \frac{{ - b - \sqrt {{b^2} - 4ac} }}{{2a}},{v_2} = \frac{{ - b + \sqrt {{b^2} - 4ac} }}{{2a}}.
\end{equation}

To find the minimal value of the $q$-th subfunction ${f_{k,q}}(\kappa )$ on $\kappa  \in \left[ {{{\widehat \eta }_{q - 1}},{{\widehat \eta }_q}} \right]$, five cases are to be considered:

\noindent $\bullet$ Case 1:  ${b^2} - 4ac \leqslant 0$ 

$\quad\;\;\text{OR}\; {b^2} - 4ac > 0{\text{ and }} {{{{v}}_1}} < {v_2} < {\widehat \eta _{q - 1}} < {\widehat \eta _q}$ 

$\quad\;\;\text{OR}\; {b^2} - 4ac > 0{\text{ and }}{\widehat \eta _{q - 1}} < {\widehat \eta _q} <  {{{v}}_1} < {v_2}$
  
${f'_{k,q}}(\kappa ) \geqslant 0$ on $[ {{{\widehat \eta }_{q - 1}},{{\widehat \eta }_q}} ]$, which means that ${f_{k,q}}(\kappa )$ is monotonically increasing on $[ {{{\widehat \eta }_{q - 1}},{{\widehat \eta }_q}} ]$. Therefore, ${f_{k,q}}(\kappa )$ has its minimal value at ${\kappa _q} = {\widehat \eta _{q - 1}}$.

\noindent $\bullet$ Case 2: ${b^2} - 4ac > 0{\text{ and }} {{{v}}_1} < {\widehat \eta _{q - 1}} < {\widehat \eta _q} < {v_2}$

${f'_{k,q}}(\kappa ) < 0$ on $[ {{{\widehat \eta }_{q - 1}},{{\widehat \eta }_q}} ]$, which means that ${f_{k,q}}(\kappa )$ is monotonically decreasing on $[ {{{\widehat \eta }_{q - 1}},{{\widehat \eta }_q}} ]$. Therefore, ${f_{k,q}}(\kappa )$ has its minimal value at ${\kappa _q} = {\widehat \eta _q}$.

\noindent $\bullet$ Case 3: ${b^2} - 4ac > 0{\text{ and }} {{{v}}_1} < {\widehat \eta _{q - 1}} < {v_2} < {\widehat \eta _q}$

${f'_{k,q}}(\kappa ) < 0{\text{ on }}[ {{{\widehat \eta }_{q - 1}},{v_2}} ]{\text{ and  }}{f'_{k,q}}(\kappa ) > 0{\text{ on }}[ {{v_2},{{\widehat \eta }_q}} ]$, which means that ${f_{k,q}}(\kappa )$ is monotonically decreasing on $[ {{{\widehat \eta }_{q - 1}},{v_2}} ]$ and monotonically increasing on $[ {{v_2},{{\widehat \eta }_q}} ]$. Therefore, ${f_{k,q}}(\kappa )$ has its minimal value at ${\kappa _q} = {v_2}$.

\noindent $\bullet$ Case 4: ${b^2} - 4ac > 0{\text{ and }}{\widehat \eta _{q - 1}} <  {{{v}}_1} < {\widehat \eta _q} < {v_2}$

${f'_{k,q}}(\kappa ) > 0{\text{ on }}[ {{{\widehat \eta }_{q - 1}},{v_1}} ]{\text{ and  }}{f'_{k,q}}(\kappa ) < 0{\text{ on }}[ {{v_1},{{\widehat \eta }_q}} ]$, which means that ${f_{k,q}}(\kappa )$ is monotonically increasing on $[ {{{\widehat \eta }_{q - 1}},{v_1}} ]$ and monotonically decreasing on $[ {{v_1},{{\widehat \eta }_q}} ]$. Therefore, ${f_{k,q}}(\kappa )$ has its minimal value at ${\kappa _q} = {\mathop {\min }\limits_{ \substack{ \kappa} }} \{ {f_{k,q}}({{\widehat \eta }_{q - 1}}),{f_{k,q}}({{\widehat \eta }_q})\}$.

\noindent $\bullet$ Case 5: ${b^2} - 4ac > 0{\text{ and }}{\widehat \eta _{q - 1}} < {{{v}}_1} < {v_2} < {\widehat \eta _q}$

${f'_{k,q}}(\kappa ) > 0{\text{ on }}[ {{{\widehat \eta }_{q - 1}},{v_1}} ]$, ${f'_{k,q}}(\kappa ) < 0{\text{ on }}[ {{v_1},{v_2}}]$ and ${\text{ }}{f'_{k,q}}(\kappa ) > 0{\text{ on }}[ {{v_2},{{\widehat \eta }_q}} ]$, which means that ${f_{k,q}}(\kappa )$ is monotonically increasing on $[ {{{\widehat \eta }_{q - 1}},{v_1}} ]$, monotonically decreasing on $[ {{v_1},{v_2}} ]$ and increasing on $[ {{v_2},{{\widehat \eta }_q}} ]$. Therefore, ${f_{k,q}}(\kappa )$ has its minimal value at ${\kappa _q} = {\mathop {\min }\limits_{ \substack{ \kappa }}}\{ {{f_{k,q}}({{\widehat \eta }_{q - 1}}),{f_{k,q}}({v_2})} \}$.

Using the method above to find the potential local minimal subfunction values ${f_{k,q}}({\kappa _q})$, $q = 1, \ldots ,Q$. Then, the minimal value for $k$-th subcarrier, $\eta _k^{\star}$, is selected from  {$Q$} local minimal values, which can be obtained:
\begin{equation}\label{45}
    \eta _k^{\star} = {\{ {\arg {\rm{ }}\mathop {\min }\limits_{{\kappa _q}} {\rm{ }}\{ {{f_{k,1}}({\kappa _1}), \ldots ,{f_{k,Q}}({\kappa _Q})} \}{\rm{ }}} \}^2}.
\end{equation}

Substituting \eqref{45} into \eqref{38}, the optimal solution ${\mathbf{h}}_{k,s}^{\star}$ can be also obtained.

\subsubsection{Optimize $\{ {{\bf{g}}_{k,m}},{\varepsilon _k}\}$ } With given other variables, $\{ {{\bf{g}}_{k,m}},{\varepsilon _k}\}$ can be updated by solving the subproblems.

\noindent
(4-i) When $i=1$: the subproblem is based on maximizing the minimum mainlobe level.
\begin{subequations}\label{46}
    \begin{align}
        &\mathop {\min }\limits_{\{ {\bf{g}}_{k,m}^{},{\varepsilon _k}\} }  - \log ({\varepsilon _k}) + \frac{{{\rho _{4,k}}}}{2}\sum\limits_{m = 1}^M {\| {{{\bf{g}}_{k,m}} - \widehat {\bf{g}}_{k,m}^{}} \|_F^2} \label{46a}\\
        &\;\;\;\;\;\text{s.t.}\;\quad\| {{{\bf{g}}_{k,m}}} \|_F^2 \ge {\varepsilon _k},{\rm{ }}\forall m,{{k}}, \label{46b}
    \end{align}
\end{subequations}
where $\widehat {\bf{g}}_{k,m}^{} = {\bf{Y}}_k^H{\bf{a}}({f_{{k}}},{\theta _m}) - {\bm{\nu }}_{k,m}^{},{\rm{ }}\forall m,{{k}}$.

It can be observed that ${\varepsilon}_k$ and $\{ {{\mathbf{g}}_{k,m}^{}} \}$ are coupled under the constraint $\| {{{\bf{g}}_{k,m}}} \|_F^2 \ge {\varepsilon _k}$. Therefore, if $\varepsilon _k^{\star}$ is known, the optimal ${\mathbf{g}}_{k,m}^{\star}$ can be determined by solving
\begin{subequations}\label{47}
    \begin{align}
        &\mathop {\min }\limits_{\{ {{\bf{g}}_{k,m}^{}} \}} \sum\limits_{m = 1}^M {\| {{{\bf{g}}_{k,m}} - \widehat {\bf{g}}_{k,m}^{}} \|_F^2} \\
        &\;\;\;\;\text{s.t.}\;\;\| {{{\bf{g}}_{k,m}}} \|_F^2 \ge \varepsilon _k^{\star},{\rm{ }}\forall m.
    \end{align}
\end{subequations}

The solution to \eqref{47} can be obtained by 
\begin{equation}\label{48}
    {\bf{g}}_{k,{\rm{m}}}^{\star} = 
    \left\{ \begin{array}{ll}
    \widehat {\bf{g}}_{k,m},&{\| {\widehat {\bf{g}}_{k,m}^{}} \|_F} > \sqrt{\varepsilon _k^{\star}}\\
    \sqrt {\varepsilon _k^{\star}} \frac{{\widehat {\bf{g}}_{k,m}^{}}}{{{{\| {\widehat {\bf{g}}_{k,m}^{}} \|}_F}}},&{\rm{ otherwise}}.
\end{array} \right.
\end{equation}

Substituting \eqref{48} into \eqref{46a}, we can transform the initial problem to the optimization problem about a function dependent on a single variable ${\varepsilon}_k$:
\begin{equation}
\!\!\mathop {\min }\limits_{{\varepsilon _k}} d_k({\varepsilon _k})\! =\!  - \!\log ({\varepsilon _k}) \!+\! \frac{{{\rho _{4,k}}}}{2}\!\!\sum\limits_{m = 1}^M \!{{\Xi _{k,m}}{{( {\sqrt {{\varepsilon _k}} \! -\! {{\| {\widehat {\bf{g}}_{k,m}^{}} \|}_F}} )}^2}} \! ,\!\!\label{49b}
\end{equation}
where ${\Xi _{k,m}} = 0$ if $\|{\widehat {\bf{g}}_{k,m}^{}} \|_F > \sqrt {{\varepsilon _k}}$, otherwise, ${\Xi _{k,m}} = 1$.

We define $[ {{{\widehat \varepsilon }_0}, \ldots ,{{\widehat \varepsilon }_Q}} ]$ be the ascending order sequence of $[ {{{\| {\widehat {\bf{g}}_{k,1}^{}} \|}_F}, \ldots ,{{\| {\widehat {\bf{g}}_{k,M}^{}} \|}_F}} ]$ without repetitive terms, and $Q \leqslant M$. Then, \eqref{49b} can be represented as a piece-wise function:
\begin{equation}
    {d_k}({\varepsilon _k}) = \{ {{d_{k,q}}({\varepsilon _k})|{{\widehat \varepsilon }_{q - 1}} \le \sqrt {{\varepsilon _k}}  \le {{\widehat \varepsilon }_q}, \forall q } \},
\end{equation}
where ${d_{k,q}}({\varepsilon}_k ) =  - \log ({\varepsilon _k}) + \frac{{{\rho _{4,k}}}}{2}\sum\nolimits_{n = 1}^q {{{( {\sqrt {{\varepsilon _k}}  - {{\widehat \varepsilon }_n}} )}^2}}$ is defined on the interval $[ {{{\widehat \varepsilon }_{q - 1}},{{\widehat \varepsilon }_q}} ]$.

Since $- \log ({\varepsilon}_k )$ is convex and $\frac{{{\rho _{4,k}}}}{2}\sum\nolimits_{n = 1}^q {{{( {\sqrt {\varepsilon}_k   - {{\widehat \varepsilon }_n}} )}^2}}$ is also convex, ${d_{k,q}}({\varepsilon}_k)$ is convex. To solve this convex problem, an auxiliary variable $\iota = \sqrt {\varepsilon}_k   > 0$ is introduced and ${d_{k,q}}({\varepsilon}_k )$ can be transformed into:
\begin{equation}
    {d_{k,q}}(\iota) =  - 2\log (\iota) + \frac{{{\rho _{4,k}}}}{2}\sum\limits_{n = 1}^q {{\iota^2}}  - \frac{{{\rho _{4,k}}}}{2}\sum\limits_{n = 1}^q {{{\widehat \varepsilon }_n}\iota}  + \frac{{{\rho _{4,k}}}}{2}\sum\limits_{n = 1}^q {{{\widehat \varepsilon }_n}^2} .
    \nonumber
\end{equation}

Similarly, to find the minimal value of ${d_{k,q}}(\iota)$, we set ${d'_{k,q}}(\iota) = 0$, which can be represented as a quadratic equation with one unknown variable:
\begin{equation}\label{52}
    \hat a{\iota^2} + \hat b\iota + \hat c = 0,
\end{equation}
where $\hat a \!=\! {\rho _{4,k}}\sum\nolimits_{n = 1}^q 1 \! >\! 0,\;\hat b \!= \! - {\rho _{4,k}}\sum\nolimits_{n = 1}^q {{{\widehat \varepsilon }_n}}  \!< \!0,\;\hat c \!= \! - 2$.

It is obvious that ${\hat b^2} - 4\hat a\hat c > 0$ and $\frac{{\hat c}}{{\hat a}} < 0$, so \eqref{52} has one non-negative root:
\begin{equation}
    {v_1} = \frac{{ - \hat b + \sqrt {{{\hat b}^2} - 4\hat a\hat c} }}{{2\hat a}}.
\end{equation}

To find the minimal value of the $q$-th subfunction   on $\iota \in [ {{{\widehat \varepsilon }_{q - 1}},{{\widehat \varepsilon }_q}} ]$, there are three cases to be considered:

\noindent $\bullet$ Case 1: ${v_1} \leqslant {\widehat \varepsilon _{q - 1}} \leqslant {\widehat \varepsilon _q}$

${d'_{k,q}}(\iota) \geqslant 0{\text{ on }}[ {{{\widehat \varepsilon }_{q - 1}},{{\widehat \varepsilon }_q}} ]$, which means that ${d_{k,q}}(\iota)$ is monotonically increasing on $[ {{{\widehat \varepsilon }_{q - 1}},{{\widehat \varepsilon }_q}} ]$. Therefore, ${d_{k,q}}(\iota)$ has its minimal value at ${\iota_q} = {\widehat \varepsilon _{q - 1}}$.

\noindent $\bullet$ Case 2: ${\widehat \varepsilon _{q - 1}} \leqslant {v_1} \leqslant {\widehat \varepsilon _q}$

${d'_{k,q}}(\iota) \leqslant 0{\text{ on }}[ {{{\widehat \varepsilon }_{q - 1}},{v_1}} ]{\text{ and }}{d'_{k,q}}(\iota) \geqslant 0{\text{ on }}[ {{v_1},{{\widehat \varepsilon }_q}} ]$, which means that ${d_{k,q}}(\iota)$ is monotonically decreasing on $[ {{{\widehat \varepsilon }_{q - 1}},{v_1}} ]$ and monotonically increasing on $[ {{v_1},{{\widehat \varepsilon }_q}} ]$. Therefore, ${d_{k,q}}(\iota)$ has its minimal value at ${\iota_q} = {v_1}$.

\noindent $\bullet$ Case 3: ${\widehat \varepsilon _{q - 1}} \leqslant {\widehat \varepsilon _q} \leqslant {v_1}$

$ {d'_{k,q}}(\iota) \leqslant 0{\text{ on }}[ {{{\widehat \varepsilon }_{q - 1}},{{\widehat \varepsilon }_q}} ]$, which means that ${d_{k,q}}(\iota)$ is monotonically decreasing on $[ {{{\widehat \varepsilon }_{q - 1}},{{\widehat \varepsilon }_q}} ]$. Therefore, ${d_{k,q}}(\iota)$ has its minimal value at ${\iota_q} = {\widehat \varepsilon _q}$.

Using the method above to find the potential local minimal subfunction values ${d_{k,q}}(\iota{}_q)$, $q = 1, \ldots ,Q$. Then, the minimal value for $k$-th subcarrier, $\varepsilon _k^{\star}$, is selected from $Q$ local minimal values, which can be given by:
\begin{equation}\label{54}
\varepsilon _k^{\star} = {\{ {\arg {\rm{ }}\mathop {{\rm{min}}}\limits_{\iota{}_q} {\rm{ }}\{ {{d_{k,1}}(\iota{}_1), \ldots ,{d_{k,Q}}(\iota{}_Q)} \}} \}^2}.
\end{equation}

Substituting \eqref{54} into \eqref{48}, the optimal solution ${\mathbf{g}}_{k,{\text{m}}}^{\star}$ can be also obtained.

\noindent
(4-ii) When $i=2$: the subproblem is based on maximizing the integrated mainlobe level.
\begin{equation}\label{IM_1}
    \mathop {\min }\limits_{\{ {\bf{g}}_{k,m}\} } - \log ( {\sum\limits_{m = 1}^M\!\! {\| {{{\bf{g}}_{k,m}}} \|_F^2} } ) + \frac{{{\rho _{4,k}}}}{2}\!\!\sum\limits_{m = 1}^M \!\! {\| {{{\bf{g}}_{k,m}} -\widehat {\bf{g}}_{k,m}^{}} \|_F^2},
\end{equation}
where $\widehat {\bf{g}}_{k,m}^{} = {\bf{Y}}_k^H{\bf{a}}({f_{{k}}},{\theta _m}) - {\bm{\nu }}_{k,m}^{},{\rm{ }}\forall m,{{k}}$.
Similar to the solving procedure of subproblem \eqref{IS_1}, we do not need to optimize ${\varepsilon _k}$ and the problem \eqref{IM_1} can be written into 
\begin{equation}\label{IM_2}
    \mathop {\min }\limits_{{{\bf{t}}_k}} \quad  - \log \| {{{\bf{t}}_k}} \|_F^2 + \frac{{{\rho _{4,k}}}}{2}\| {{{\bf{t}}_k} - {{{\bf{\hat t}}}_k}} \|_F^2,
\end{equation}
where ${{\bf{t}}_k} = [{{\bf{g}}_{k,1}}; \ldots ;{{\bf{g}}_{k,M}}]$ and ${{{\bf{\hat t}}}_k} = [{{{\bf{\hat g}}}_{k,1}};\ldots ;{{{\bf{\hat g}}}_{k,M}}]$.

Similarly, by taking the first-order Taylor expansion of the first term in \eqref{IM_2} with respect to ${{\bf{t}}_k}$ at its last iteration point ${{{{\bf{\bar t}}}_k}}$, the objective function in \eqref{IM_2} is converted into
\begin{equation}\label{IM_3}
    \mathop {\min }\limits_{{{\bf{t}}_k}} \; - {\log \| {{{{\bf{\bar t}}}_k}} \|_F^2 \! +\! \frac{{{{{\bf{\bar t}}}_k}}}{{\| {{{{\bf{\bar t}}}_k}} \|_F^2}}( {{{\bf{t}}_k} \!-\! {{{\bf{\bar t}}}_k}} )}\! +\! \frac{{{\rho _{4,k}}}}{2}\| {{{\bf{t}}_k}\! -\! {{{\bf{\hat t}}}_k}} \|_F^2,
\end{equation}
The optimal solution to \eqref{IM_3} can be easily derived as ${{\bf{t}}_k^{\star}} = {{{\bf{\hat t}}}_k} + {{{{{\bf{\bar t}}}_k}}}/({{{\rho _{4,k}}\| {{{{\bf{\bar t}}}_k}} \|_F^2}})$.
Accordingly, the optimal ${\mathbf{g}}_{k,{\text{m}}}^{\star}$ can be also obtained.

\subsubsection{Optimize ${{\bf{F}}_k}$} 
With given other variables, ${\bf{F}}_{k} $ can be updated by solving the subproblem:
 \begin{equation}
     \mathop {\min }\limits_{{{\bf{F}}_k}} \;\| {{{\bf{V}}_k} - {{\bf{F}}_\mathrm{RF}}{{\bf{F}}_k}} \|_F^2 ,
     \label{26}
 \end{equation}
where ${{\mathbf{V}}_k} = \left[ {{{\mathbf{y}}_{k,1}} + {{\bm{\varsigma }}_{k,1}}, \ldots ,{{\mathbf{y}}_{k,U}} + {{\bm{\varsigma }}_{k,U}}} \right] \in {\mathbb{C}^{{M_t} \times U}}$. The optimal solution of \eqref{26} can be derived via the first-order condition as
\begin{equation}
    {\bf{F}}_k = {\left( {{\bf{F}}_{\mathrm{RF}}^H{{\bf{F}}_{\mathrm{RF}}}} \right)^{ - 1}}{\bf{F}}_{\mathrm{RF}}^H{{\bf{V}}_k} .
    \label{27}
 \end{equation}
 
 \subsubsection{Optimize ${{\bf{F}}_{\rm{RF}}}$} 
 With given other variables, ${\bf{F}}_\mathrm{RF} $ can be updated by solving the subproblem:
\begin{subequations}
    \begin{align}
          &\mathop {\min }\limits_{{{\bf{F}}_{\text{RF}}}} {\| {{{\bf{V}}_{{k}}} - {{\bf{F}}_{\text{RF}}}{{\bf{F}}_k}} \|_F^2}  \label{28a}\\
          &{\text{ s.t. }}| {{{\mathbf{F}}_{\text{RF}}}(i,j)} | = 1,{\text{ }}\forall  i, j  . \label{28b}
    \end{align}
    \label{28}%
\end{subequations}
Generally speaking, the problem \eqref{28} is hard to tackle due to the discrete constraint \eqref{28b}.
To effectively solve it, we propose a cyclic coordinate descent (CCD) framework to tackle this problem \eqref{28}.
Specifically, the methodologies based on CCD generally start with a feasible initial matrix ${\bf F}_{{\rm RF}} = {\bf F}_{{\rm RF}}^{[0]}$.
Then, an element in ${\bf F}_{{\rm RF}} ( i,j)$ is set as the variable while others are held fixed.
The objective function is optimized with respect to this identified element.
All elements are updated one-by-one for several times until reaching the convergence condition.
\begin{prop}\label{prop_2}
With other variables being fixed, the selected element ${\bf{F}}_{\rm RF} {(i,j)}  = {e^{\jmath \gamma_{i,j}}}$  can be updated by
\begin{subequations}\label{61}
    \begin{align}
        \mathop {\min }\limits_{{{\mathbf{F}}_{\text{RF}}}(i,j)} \;\;&  2 \Re \left\{ {{{\mathbf{F}}_{\text{RF}}}(i,j){\psi _{i,j}}} \right\} + \text{const.}\label{61A}\\
        {\text{s.t. }} \quad & {{\mathbf{F}}_{\text{RF}}}(i,j) = {e^{ - \jmath{\gamma _{i,j}}}},{\gamma _{i,j}} \in (0,2\pi ],\label{61B}
    \end{align}
\end{subequations}
where ${\psi _{i,j}} =  {{{\mathbf{F}}_k}(j,:)( {\mathbf{F}}_k^H {\mathbf{T}}_{\text{RF},ij}^H(i,:) - {{\mathbf{V}}_{k}^H(i,:) })}$ and ${\mathbf{T}}_{\text{RF},ij}^{}$ is constructed to be ${{\mathbf{F}}_{\text{RF}}}$ whose $(i,j)$-th entry is zero.
Then, the closed-form solution for ${\mathbf{F}}_{\text{RF}}(i,j)$ can be derived as
\begin{equation}
    {\mathbf{F}}_{\text{RF}}(i,j) = {\mathrm{exp} \left\{ - \jmath (\angle {\psi _{i,j}} + \pi) \right\}}
\end{equation}
\end{prop}
\begin{IEEEproof}
    Please refer to Appendix \ref{app1}.
\end{IEEEproof}
According to descriptions above, the analog beamformer design algorithm is completed and summarized in Algorithm \ref{alg1}.

\setlength{\textfloatsep}{0em}
\begin{algorithm}[!t]
	\caption{CCD-Based Method to Solve the Problem \eqref{28}}
	\label{alg1}
	\begin{algorithmic}[1]
		\STATE \textbf{Input:} ${\bf{F}}_{\rm{RF}}^t$, $N_{\text{CCD}}^{\max }$.
		\STATE Set ${\bf{F}}_{\rm{RF}}^{[0]} = {\bf{F}}_{\rm{RF}}^{t} $, $l=0$.
        \REPEAT
            \STATE $l = l + 1$
            \FOR{$i=1,...,M_t$}
                \FOR{$ j=1,...,N_t$}
                    % \STATE  Compute ${\psi _{i,j}}$ and $\gamma _{i,j}^{\star} = \angle {\psi _{i,j}} + \pi$.
                    \STATE  Update ${\mathbf{F}}_{\text{RF}}^{[l]}(i,j) = {\mathrm{exp} \left\{ - \jmath (\angle {\psi _{i,j}} + \pi) \right\}}$.
                \ENDFOR
            \ENDFOR 
        \UNTIL  $l = N_{\text{CCD}}^{\max }$ or \textit{Convergence}.
		\STATE  Return ${\bf{F}}_{\rm{RF}}^{t+1} = {\bf{F}}_{\rm{RF}}^{[l]}$.
	\end{algorithmic}
\end{algorithm}

\subsubsection{Update $( {{\bm{\varsigma }}_{k,u}} , \lambda _{k,u}^{k,i} , {{\bm{\beta }}_{k,s}}, {{\bm{\nu }}_{k,m}} )$} Four dual variables can be updated by doing the following operations\footnote{The notation $t$ denotes the iteration number.} \cite{boyd2011distributed}:
\begin{subequations}
    \begin{align}
        &{\bm{\varsigma }}{_{k,u}^{t + 1}} = { \bm{\varsigma }_{k,u}^t} +  {\bf{y}}_{k,u}^{t + 1} - {\bm{b}}_{k,u}^{t + 1} \\
        &{\{ \lambda _{k,u}^{k,i}\} ^{t + 1}} \!=\! {\{ \lambda _{k,u}^{k,i}\} ^t}\! +\! ({\{ G_{k,u}^{k,i}\} ^{t + 1}} \!-\! {\bf{w}}_{k,u}^H{{\bf{H}}_{k,u}}{\bf{y}}_{k,i}^{t + 1})\\
        &{\bm{\beta }}{_{k,s}^{t + 1}} = {\bm{\beta }}{_{k,s}^t} + ({\bf{h}}_{k,s}^{t + 1} - { \{{\bf{Y}}_k^{t + 1}\} ^H}{\bf{a}}({f_{{k}}},{\vartheta _s}))\\
        & {\bm{\nu }}_{k,m}^{t + 1}=  {\bm{\nu }_{k,m}^t} + ({\bf{g}}_{k,m}^{t + 1} - {\{ {\bf{Y}}_k^{t + 1}\} ^H}{\bf{a}}({f_{{k}}},{\theta _m})).
    \end{align}
    \label{55}%
\end{subequations}
For clarity, the procedure of the proposed task-oriented HBF design algorithm in a unified CADMM framework can be summarized in Algorithm \ref{alg2}.

\setlength{\textfloatsep}{0.5em}
\begin{algorithm}[!t]
	\caption{The Task-Oriented Transmit HBF Algorithm}
	\label{alg2} {
	\begin{algorithmic}[1]
		\STATE \textbf{Input:} $U$, $K$, $M_t$, $N_t$, $M_r$, $B$, $f_c$, ${\bf{\Theta }}$, ${\bf{\Omega }}$, ${\bf{\Phi }}$, penalty parameters $ {\rho}_{1,k},~\rho_{2,k},~\rho_{3,k} ,~\rho_{4,k} >0 $, task index $i$, and the maximum iteration number $N_{\text{CADMM}}^{\max }$.
		\STATE \textbf{Output:} $\left\{ {{\bf{F}}_k} \right\}_{k = 1}^K$ and ${\bf{F}}_{\text{RF}}$.
		\STATE \textbf{Initialization:} set $t=0$, initialize $\left\{ {{{\bf{F}}_k}} \right\}_{k = 1}^K$ and ${{\bf{F}}_{\text{RF}}}$.
		\REPEAT
		\STATE  \textbf{(I)} Update $\{{\mathbf{Y}}_{{k}}\}^{t+1}$ by solving \eqref{21} with the aid of the bisection method finding $\mu _k^{\star}$.
        \STATE  \textbf{(II)} Update ${{{\{ G_{k,u}^{k,i}\} }^{t + 1}}}$ by solving problem \eqref{33} with the aid of the bisection method finding ${\varphi ^{\star}}$.
        \IF{\{$i=1$\}} 
        \STATE \textbf{(III-i)} \textbf{AISMMR-based design approach:}\\
        \vspace{-1em}
        \begin{equation}
        \left
        \{\begin{aligned}
        &{\text{Compute}}\; \{ {\bm{g}}_{k,m},\varepsilon _k\}^{t + 1}\;{\text{by}}\;\eqref{48} \;{\text{and}}\;\eqref{54}.\\
        &{\text{Compute}}\; \{ {\bm{h}}_{k,s}\}^{t + 1}\;{\text{by}}\;\eqref{IS_1}.
        \end{aligned}\nonumber
        \right.
        \end{equation}
        \vspace{-0.5em}
        \ELSE[$i=2$]
        \STATE \textbf{(III-ii)} \textbf{APSIMR-based design approach:}\\
        \vspace{-1em}
        \begin{equation}
        \left
        \{\begin{aligned}
        &{\text{Compute}}\; \{ {\bm{h}}_{k,s},{{\eta }}_k\}^{t + 1}\;{\text{by}}\;\eqref{38} \;{\text{and}}\;\eqref{45}.\\
        &{\text{Compute}}\; \{ {\bm{g}}_{k,m}\}^{t + 1}\;{\text{by}}\;\eqref{IM_1}.
        \end{aligned}\nonumber
        \right.
        \end{equation}
        \vspace{-0.5em}
        \ENDIF 
        \STATE  \textbf{(IV)} Update $\{{\mathbf{F}}_{{k}}\}^{t + 1}$ by \eqref{27}.
		\STATE  \textbf{(V)} Update ${\mathbf{F}}_{{\text{RF}}}^{t + 1}$ by Algorithm \ref{alg1}.
		\STATE  \textbf{(VI)} Update dual variables by \eqref{55}.
		\STATE  set $ t=t+1 $.
        \UNTIL  $t = N_{\text{CADMM}}^{\max }$ or \textit{Convergence}.
		\STATE  Return $\{{\mathbf{F}}_{{k}}\}^{t}$ and ${\mathbf{F}}_{{\text{RF}}}^{t}$.
	\end{algorithmic}}
\end{algorithm}

\vspace{-1em}
\subsection{Complexity Analysis}
\vspace{-0.3em}

In this section, we give a brief complexity analysis of the proposed task-oriented HBF algorithm for the OFDM-DFRC systems.
Updating ${\mathbf{Y}}_k$ by solving \eqref{22} needs a complexity of ${\cal O}\left(  M_t^3 \right)$. 
Updating ${\bf{F}}_k$ by \eqref{26} needs a complexity of ${\cal O}\left(  N_t^3\right)$ and updating ${\bf{F}}_{\text{RF}}$ needs a complexity of ${\cal O}\left(  N_{\text{CCD}}{M_t}N_t^2U\right)$, where $N_{\text{CCD}}$ is the iteration number of the CCD method. 
Besides, updating ${\{ {G_{k,u}^{k,i}} \}^{t + 1}}$ by computing $\varphi$ needs a complexity of ${\cal O}\left( {M_t}{N_t}U^2\right)$. 
Solving \eqref{36} and \eqref{IS_3} needs a complexity of ${\cal O}\left( SM_tU\right)$. 
Solving \eqref{46} and \eqref{IM_3} needs a complexity of ${\cal O}\left( MM_tU\right)$.
As a result, the overall complexity of the proposed task-oriented HBF algorithm is ${\cal O} (N_{\text{CADMM}}({M_t}^3 + {N_t}^3  + N_{\text{CCD}}M_t{N_t}^2U + M_tM_rU^2 + (M+S)M_tU))$, where $N_{\text{CADMM}}$ is the iteration number needed in the CADMM.

\vspace{-0.5em}
\section{Simulation Examples}
% \vspace{-0.3em}

In this section, we evaluate the performance of the proposed HBF design for the DFRC system with multiple tasks, namely the radar scanning and detection (SD) task and the target tracking (TT) task.
Particularly, we first present a single-carrier scenario, and then examine an OFDM scenario.

\vspace{-0.5em}
\subsection{Benchmark and Desired Beampattern Setting}
\vspace{-0.3em}

For comparison, some benchmarks are considered:
1) FD-BF: the DFRC BS with APSIMR/AISMMR-based fully-digital beamformer (FD-BF) is included as the upper bound of the DFRC BS with the proposed HBF.
2) RadarOnly HBF: the DFRC BS with HBF only completes radar work.
3) ISMR-based HBF: the DFRC BS with the existing ISMR-based HBF, which is used to show the superiority of the proposed HBF.

For evaluation, we consider two different desired transmit beampatterns:
1) Desired Beampattern A: the mainlobe region ${\bf{\Theta }} = [ { - {{5}^ \circ }, {{5}^ \circ }} ]$ and sidelobe region ${\bf{\Omega }} = [ { - {{90}^ \circ }, - {{8}^ \circ }} ] \cup [ {{{8}^ \circ },{{90}^ \circ }}]$ across all the subcarriers; 
2) Desired Beampattern B: the mainlobe region ${\bf{\Theta }} = [ { - {{10}^ \circ }, {{10}^ \circ }} ]$ and  sidelobe region ${\bf{\Omega }} = [ { - {{90}^ \circ }, - {{13}^ \circ }} ] \cup [ {{{13}^ \circ },{{90}^ \circ }}]$ across all the subcarriers.

\vspace{-0.5em}
\subsection{HBF for Single-Carrier Scenario}
% \vspace{-0.3em}

\subsubsection{Simulation Setup}

In the first example, we consider a single-carrier scenario, i.e., $K=1$, unless specified otherwise.
We consider the transmitter with $M_t=32$ transmit antennas and $N_t=4$ RF chains, serving $U=4$ users.
The system transmits $N_s=4$ data streams and each user is equipped with $M_r=4$ receive antennas.
Besides, the data rate threshold of each user on each subcarrier is set identical, i.e., $\chi _{k,u} = \chi, \forall k,u$ (bits/s/Hz), and the iteration number is $N_{\text{CADMM}}=1000$.
The sampling points for the spatial angular range $\left[ { - {{90}^ \circ },{{90}^ \circ }} \right]$ is $P = 361$.
We assume that the noise variance is $\sigma _n^2 = 0$dB.
Both transmitter and receiver utilize ULAs, whose element spacing is determined by the half-wavelength.

% \vspace{-1em}
\subsubsection{Convergence of Proposed Algorithm}
Fig. \ref{fig.1-1} plots the convergence performance of the proposed task-oriented HBF algorithm for the SD task and TT task when setting Desired Beampattern B and data rate threshold $\chi\!=\!1$.
It can be observed that, in both tasks, the curves for the objective values, show a downward trend and finally reach a stationary value, while all the residual errors tend to be zero.
This validates the convergency of the proposed algorithm.
Additionally, Fig. \ref{fig.1-1-2} and Fig. \ref{fig.1-1-1} show the convergence of the achievable data rate versus the iteration number. 
The result shows that the achievable rate tends to the user-defined threshold\footnote{ {In this paper, we compare the radar performance by fixing the communication QoS threshold.}}. 
This verifies the effectiveness of the proposed algorithm, which is able to satisfy the communication QoS requirement.

\begin{figure}[t]
	\centering  
	\subfigure[]{
			\label{fig.1-1-2}
	\includegraphics[width=0.95\linewidth]{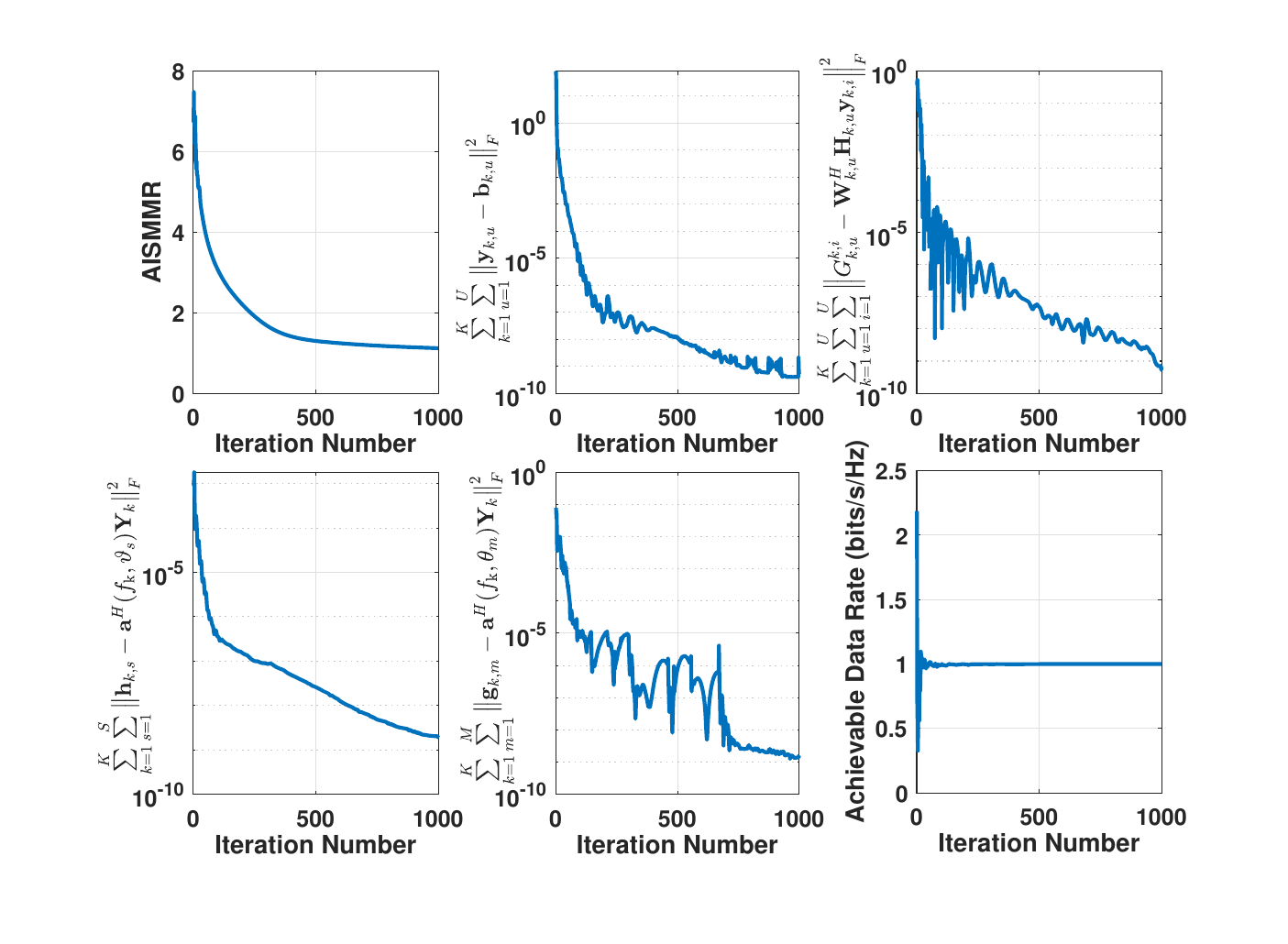}
		}\noindent
    \vspace{-0.5em}	
        \centering
        \subfigure[]{
			\label{fig.1-1-1}
    \includegraphics[width=0.95\linewidth]{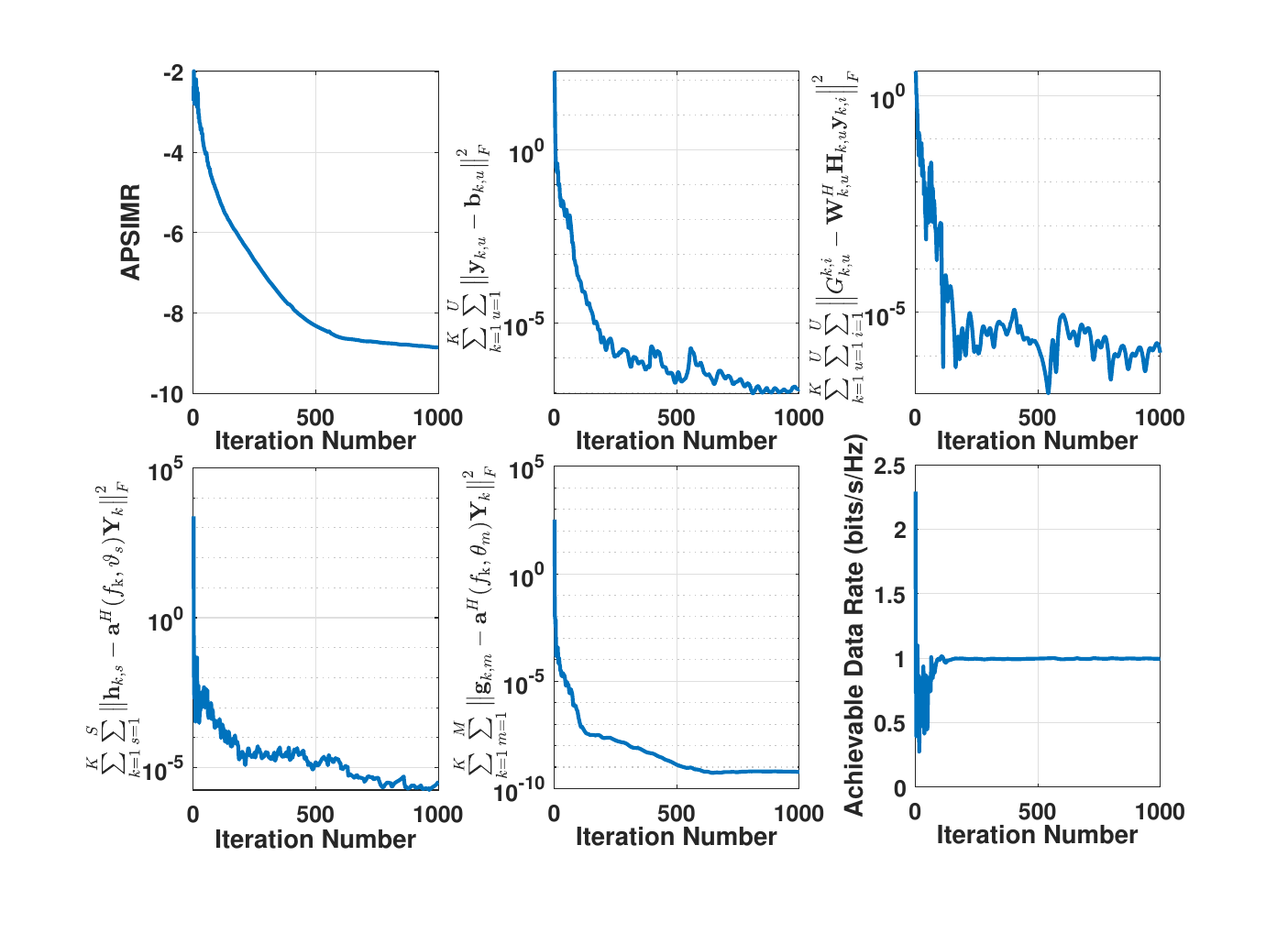}	
	}
      \vspace{-1em}
	\caption{ {Convergence performance of the proposed HBF algorithm with Desired Beampattern B and rate threshold $\chi = 1$. (a) SD task: AISMMR, residual errors, and achievable data rate versus the iteration number. (b) TT task: APSIMR, residual errors, and achievable data rate versus the iteration number.}}
	\label{fig.1-1}
 	% \vspace{-1em}
\end{figure}

Fig. \ref{fig.1-2} presents the curves of the objective values for the SD task and TT task respectively, versus the iteration number with different communication rate thresholds $\chi=1,1.5,2,2.5$, by setting Desired Beampattern A. 
Obviously, all the curves of the objective values decrease with the iterations. 
Moreover, we find that the smaller the communication rate threshold is, the lower the radar APSIMR and AISMMR can be achieved.

\begin{figure}[!ht]
    \centering
    \includegraphics[width=0.95\linewidth]{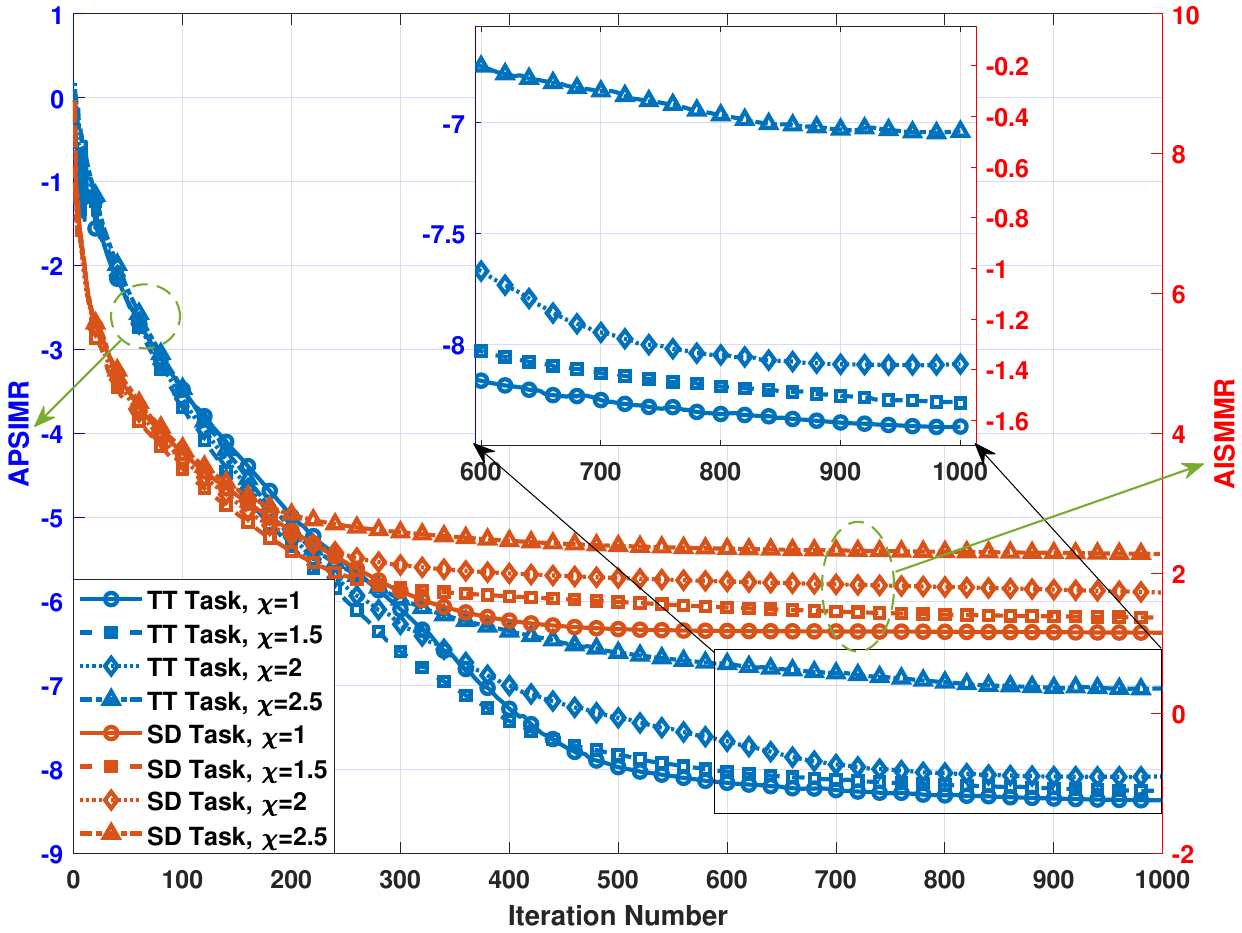}
    \vspace{-1em}
    \caption{ {Convergence comparisons of the objective functions in SD task and TT task versus the iteration number with Desired Beampattern A, setting different communication rate threshold $\chi=1,1.5,2,2.5$.}}
    \label{fig.1-2}
        % \vspace{-0.5em}
\end{figure}

\subsubsection{Radar Beampattern}

\begin{figure}[!ht]
    \centering
    \vspace{-1.2em}
    \includegraphics[width=0.95\linewidth]{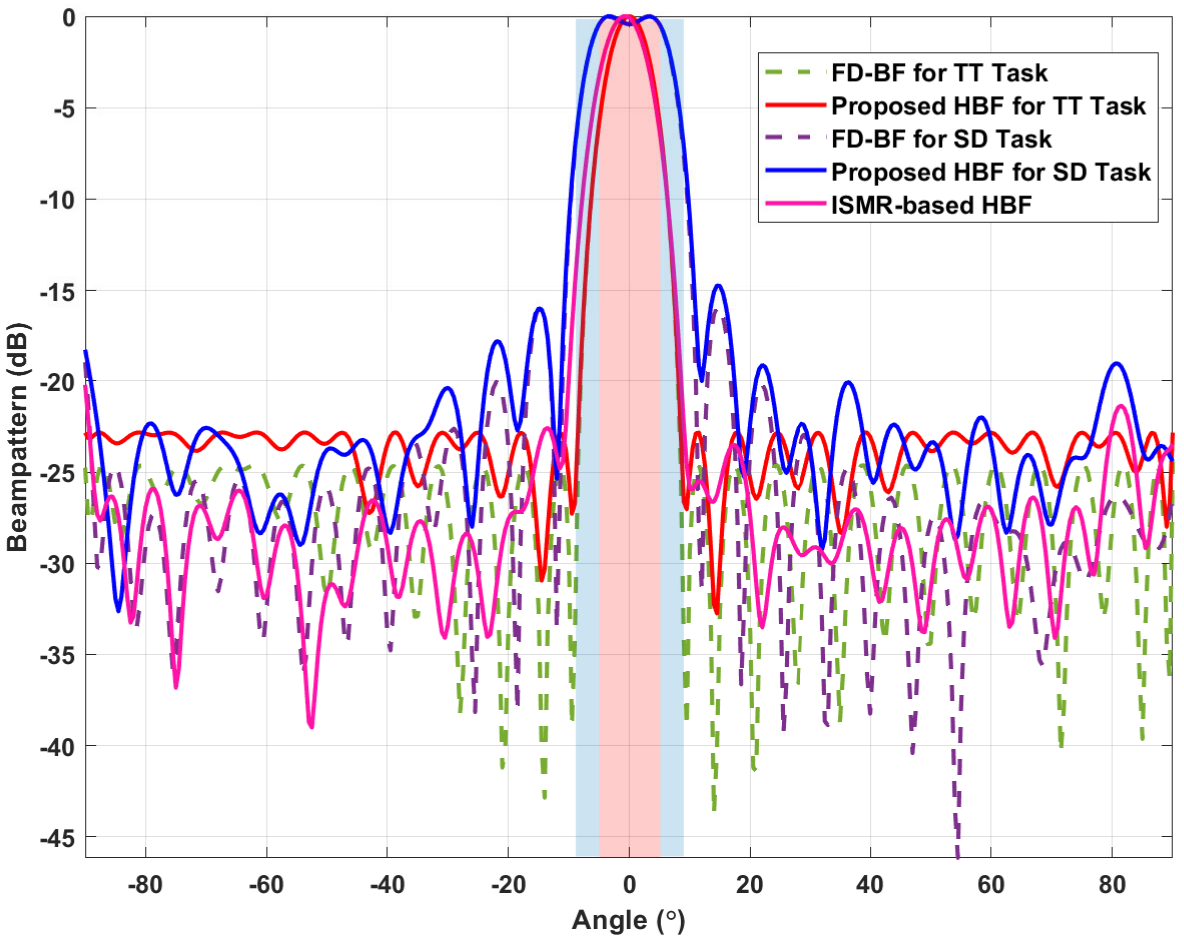}
    \vspace{-1em}
    \caption{ {Normalized radar beampatterns obtained by using different beamforming methods, with Desired Beampattern A, and rate threshold $\chi = 2$. (The desired mainlobe is in red while the desired transition band is in blue.)}}
    \label{fig.1-3}
\end{figure} 

Fig. \ref{fig.1-3} compares the normalized radar beampatterns obtained by the proposed HBF with other methods by setting Desired Beampattern A and communication rate threshold $\chi = 2$. 
As can be seen from this figure, for both tasks, the beampattern performance of the proposed HBF approaches that of the FD-BF.
Moreover, the proposed HBF for the SD task maintains the level on the whole mainlobe region while the beampattern of the ISMR-based HBF has a sharp level decrease in the mainlobe away from the $0^{\circ}$.
Besides, the proposed HBF for the TT task suppresses the sidelobe to a low level on the whole sidelobe region while some high sidelobe levels exist in the beamppatern obtained by the ISMR-based HBF.
This shows that the proposed task-oriented HBF has better beampattern properties than the simple ISMR-based HBF in both the SD task and TT task.
Therefore, it demonstrates the effectiveness and superiority of the proposed task-oriented HBF.

\begin{figure}[!ht]
    \centering
    \includegraphics[width=0.95\linewidth]{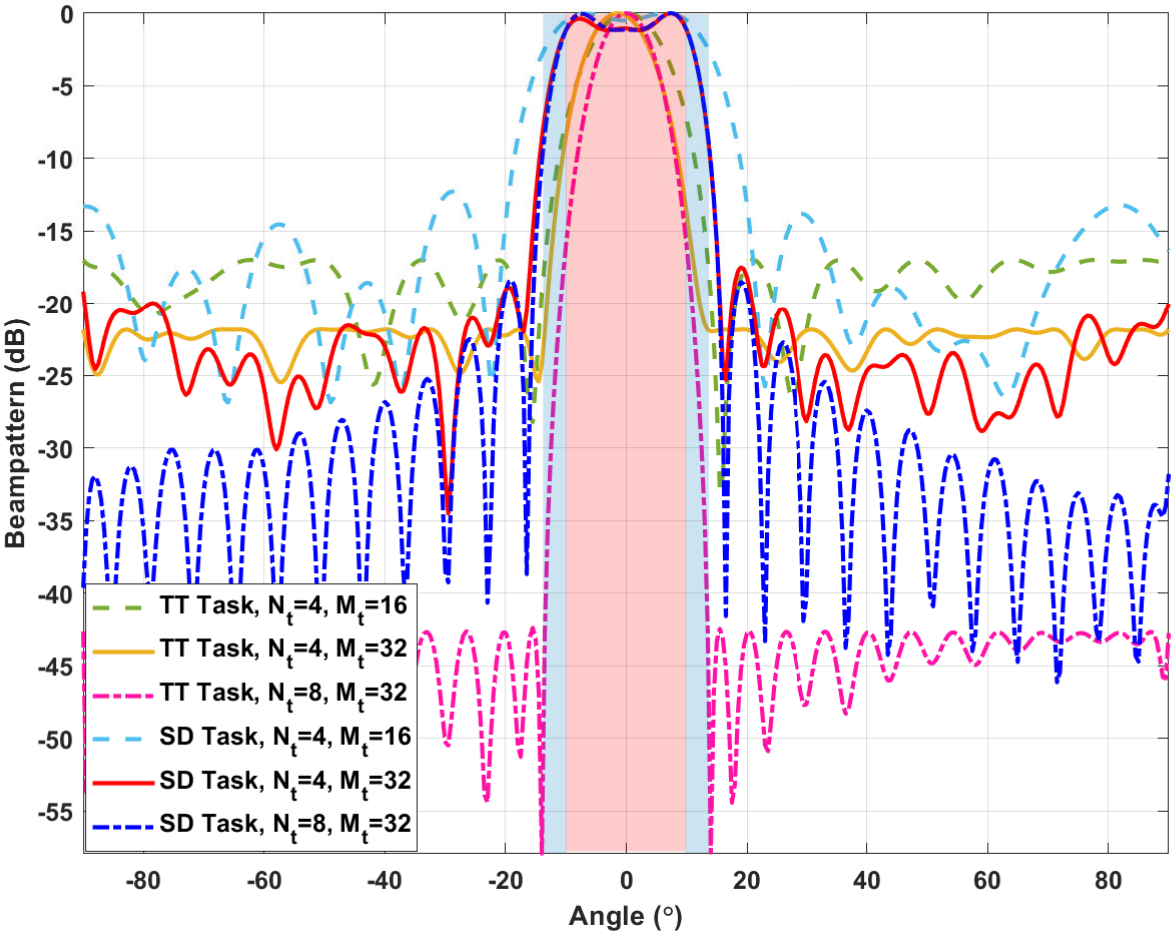}
    \vspace{-1em}
    \caption{ {Normalized radar beampatterns designed for the SD task and TT task, with Desired Beampattern B, using a different number of transmit antennas $M_t$ and RF chains $N_t$. (The desired mainlobe is in red while the desired transition band is in blue.)}}
    \label{fig.1-4}
    % \vspace{-1em}
\end{figure}

Fig. \ref{fig.1-4} plots the normalized radar beampatterns obtained by the proposed HBF with different numbers of transmit antennas $M_t=16,32$ and RF chains $N_t = 4,8$. 
We observe that the sidelobe level corresponding to $(N_t=4, M_t=16)$ is the highest for both task scenarios, and the sidelobe level is better suppressed when increasing $M_t$ and $N_t$. 
It indicates that a suitable choice of the number of transmit antennas and RF chains can enhance the radar beampattern behavior.

\begin{figure}[!ht]
\vspace{-0.5em}
    \centering
    \includegraphics[width=0.95\linewidth]{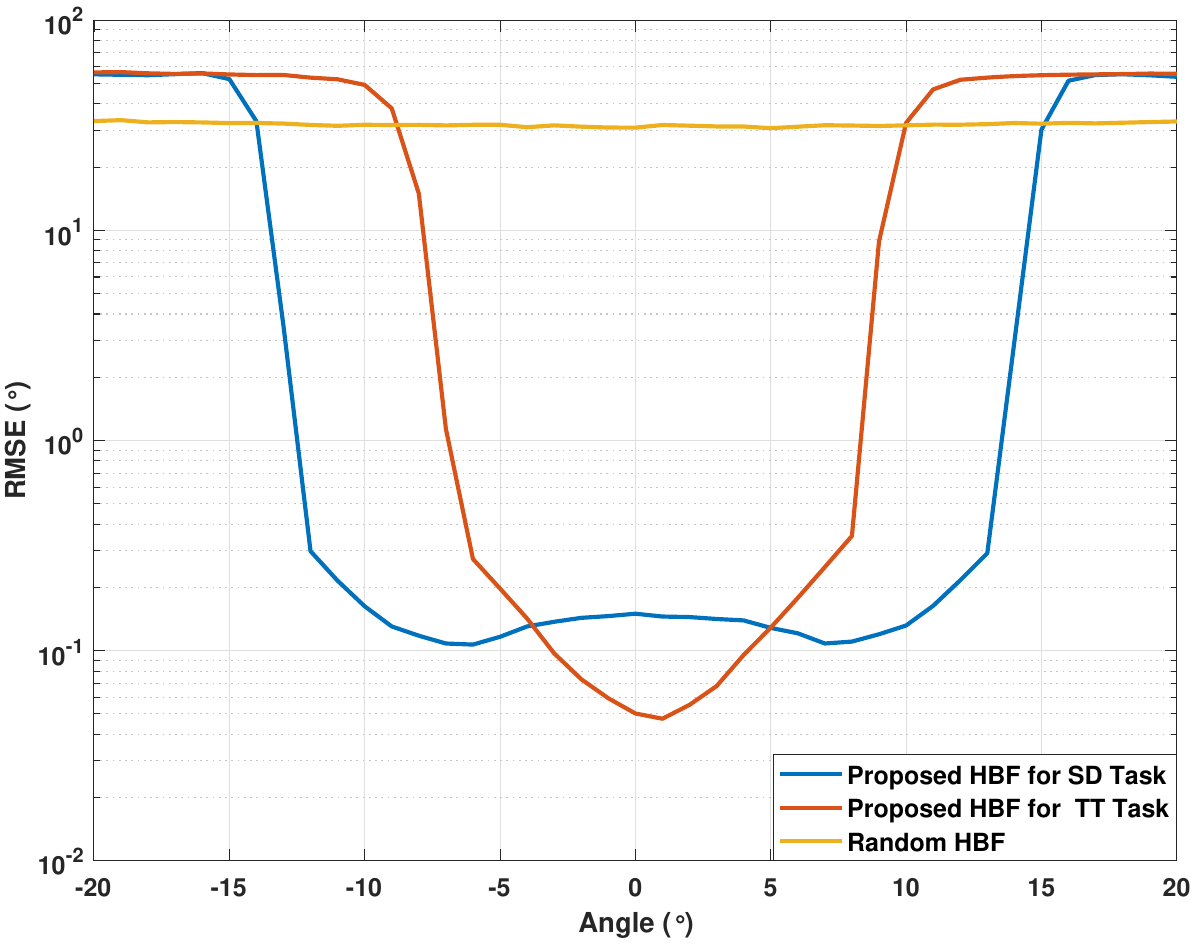}
    \vspace{-1em}
    \caption{Comparison of DOA estimation performance: the RMSEs of MUSIC-based DOA estimation versus angles.}
    \label{fig.1-5}
\end{figure} 

\subsubsection{MUSIC-based DOA Estimation Performance}
Then, we compare the MUSIC-based DOA estimation performance of the designed HBF for different modes.
The RMSE \cite{van2004detection} is defined as 
\begin{equation}
    {\text{RMSE = }}\sqrt {\frac{1}{\mathcal{N}}\sum\limits_{n \in \mathcal{N}} {{{\left( {{{\hat \theta }_n} - {\theta _n}} \right)}^2}} } \nonumber
\end{equation}
where $\mathcal{N}$ denotes Monter-Carlo trials for each angular point. ${{\theta _n}}$ and ${{{\hat \theta }_n}}$ respectively denote the actual and estimated DOA of the target. 
The radar SNR is assumed to be 10 dB, 
Fig. \ref{fig.1-5} plots the RMSEs of MUSIC-based estimation versus angles.
It can be observed that the proposed HBF for the SD task can realize relatively lower RMSEs among the whole mainlobe region than random HBF and the proposed HBF for the TT task can realize significantly lower RMSEs among a narrower mainlobe region around $0^{\circ}$ than the proposed HBF for the SD task.
Although the DOA estimation in the SD task is less accurate, it can estimate the target location over a wider region, which is more suitable for scanning.
Then, after attaining the prior information, the mode is switched from SD to TT, which can realize better DOA estimation accuracy.
Therefore, it proves that the proposed task-oriented HBF can meet different DOA estimation requirements in different task scenarios.

\vspace{-0.5em}
\subsection{Hybrid beamforming for the OFDM Scenario}
\vspace{-0.3em}

\subsubsection{Simulation Setup}
In this simulation example, we consider the HBF design for the wideband OFDM-DFRC systems.
Unless specified otherwise, the transmitter is equipped with $M_t=32$ antennas and $N_t=4$ RF chains, serving $U=4$ users.
The number of subcarriers in the OFDM-DFRC system is $K = 32$ and the center frequency is ${f_c} = 10$ GHz.
Besides, the system transmits $N_s = 4$ data streams per subcarrier.
The system bandwidth is $B = 2.56$ GHz, so the subcarrier spacing can be calculated as $\Delta f = \frac{B}{K} = 20$ MHz. 
Its percent bandwidth can be calculated as $B\%  = B/{f_c} = 25.6\%  \ge 1\%$, which indicates that this OFDM-DFRC system is a wideband system. 
The symbol duration is $\Delta t = \frac{1}{{\Delta f}} = 50$ ns. The ULAs are adopted, whose element spacing is $d = \frac{\lambda }{2} = \frac{1}{2} \cdot \frac{c}{{{f_{\max }}}} = \frac{1}{2} \cdot \frac{c}{{{f_c} + \frac{B}{2}}} = 0.0133$ m.
The target communication rate for each user at each subcarrier is identical, i.e., ${\chi _{k,u}}= {\chi},\forall k,u$.
The sampling points for spatial angular range $\left[ { - {{90}^ \circ },{{90}^ \circ }} \right]$ is $P = 361$.
The noise variance is assumed to be $\sigma _n^2 = 0$dB.
In addition, the maximum iteration number $N_{\text{CADMM}}=1000$.

\subsubsection{Convergence of Proposed Algorithm}
Fig. \ref{fig.2-1} illustrates the convergence of objective function for the SD task and TT task versus the iteration number with different rate thresholds $\chi=1.5,2,2.5,3,3.5$, setting Desired Beampattern A. 
It can be noticed that all the curves decrease at first and tend to be stable at last. 
With the increase of the communication rate threshold, both APSIMR and AISMMR become higher, which shows that different QoS constraints have different influences on space-frequency behaviors.
This verifies the effectiveness of our proposed HBF algorithm and shows that radar competes with communication.

\begin{figure}[!t]
    \centering
    \includegraphics[width=0.95\linewidth]{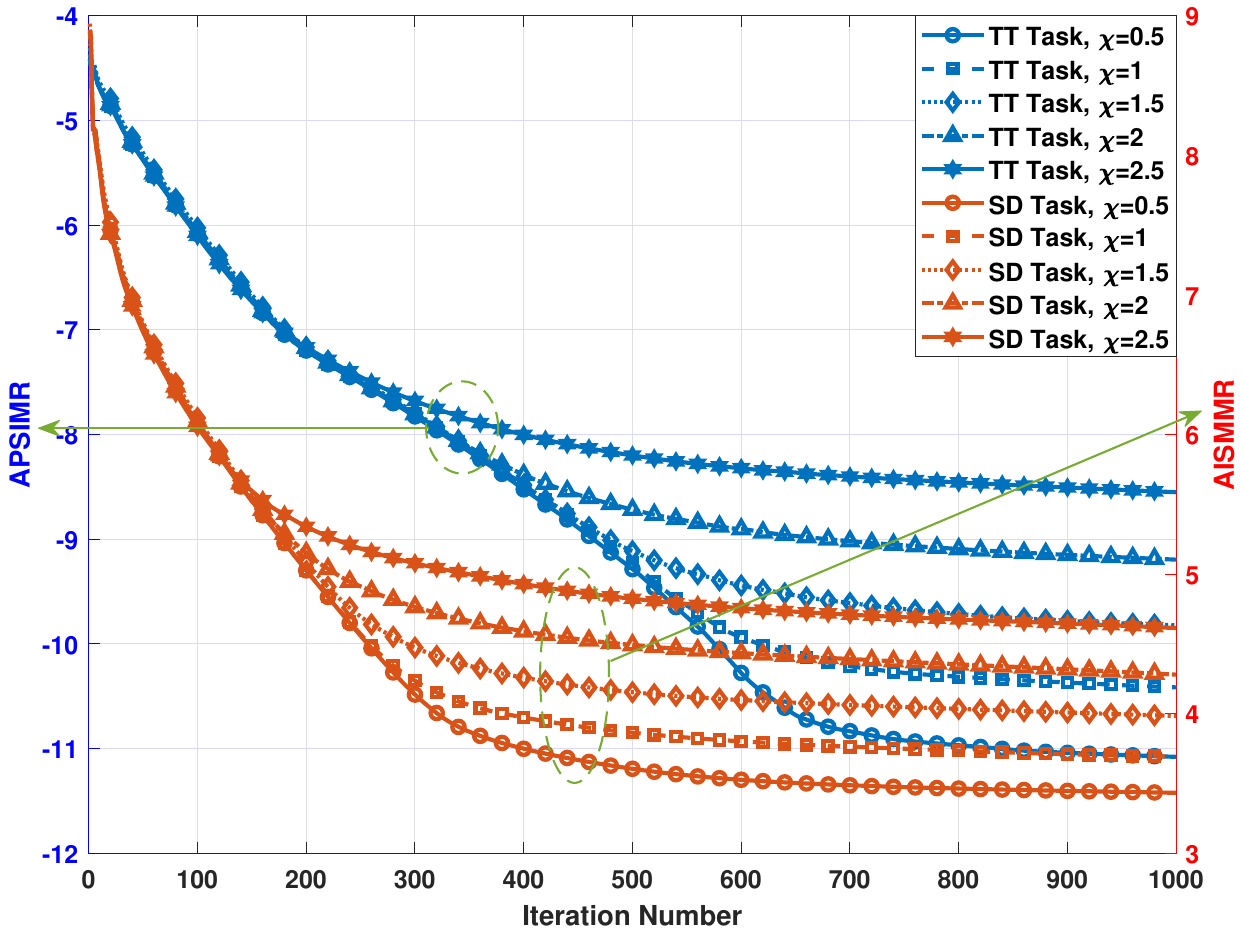}
    \vspace{-1em}
    \caption{ {Convergence comparisons of the objective functions for the SD task and TT task versus the iteration number, with Desired Beampattern A, different communication rate threshold $\chi=0.5,1,1.5,2,2.5$.}}
    \label{fig.2-1}
\end{figure}

\begin{figure*}[ht!]
\centering
\subfigure[]{
\includegraphics[width=0.327\linewidth]
{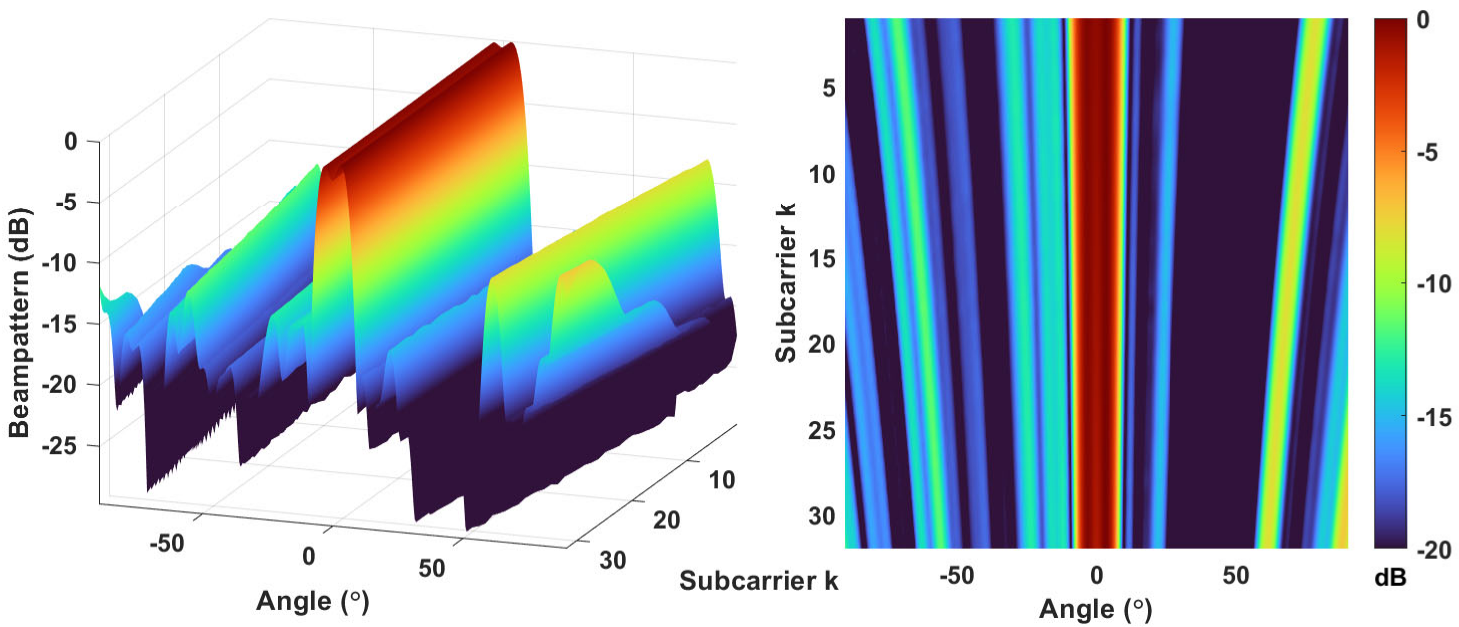} 
\label{fig.2-3-2}
}\hspace{-0.9em}
\subfigure[]{
\includegraphics[width=0.327\linewidth]
{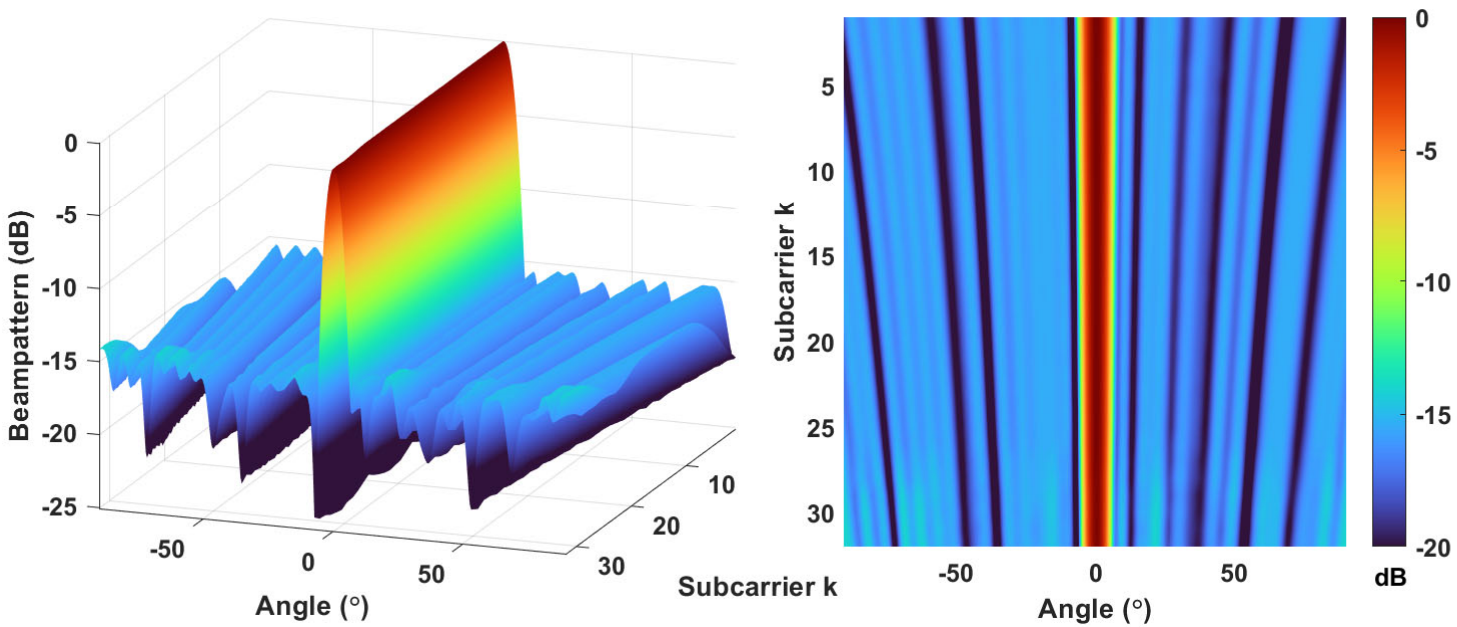} 
\label{fig.2-3-1}
}\hspace{-0.9em}
\subfigure[]{
\includegraphics[width=0.327\linewidth]{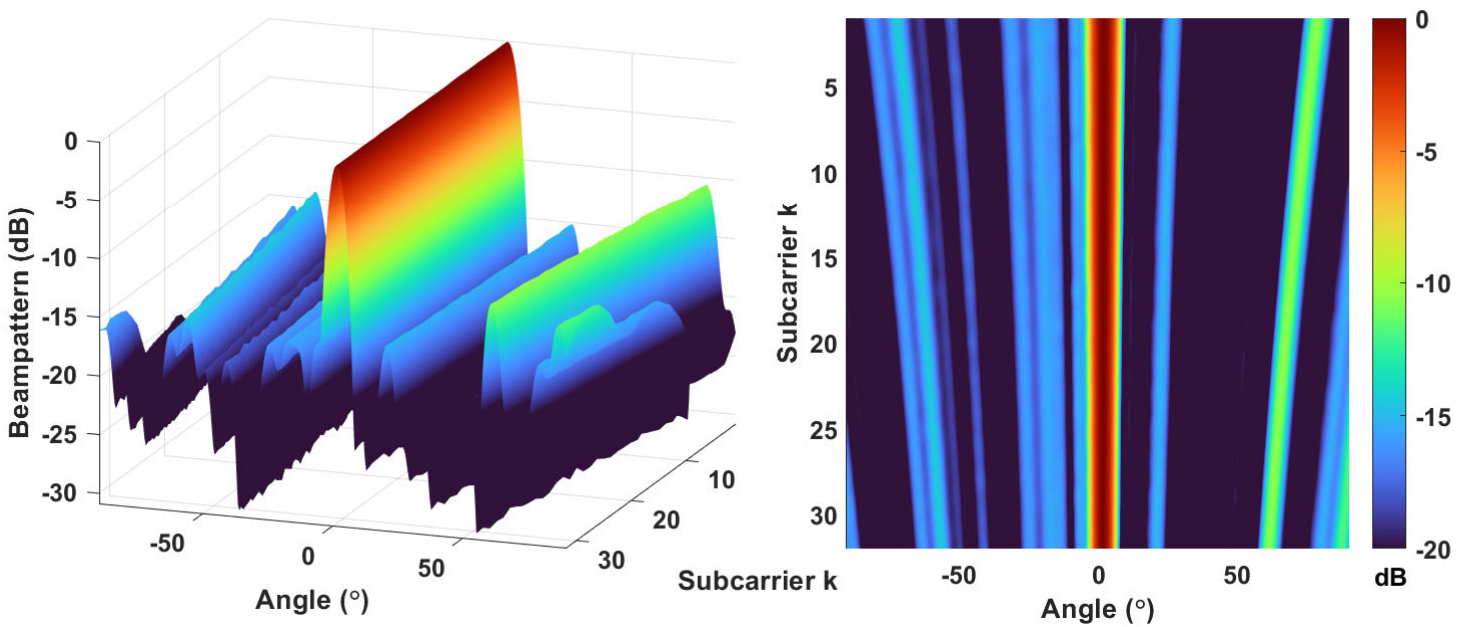}
\label{fig.2-3-3}
}\vspace{-1em}

\subfigure[]{
\includegraphics[width=0.327\linewidth]
{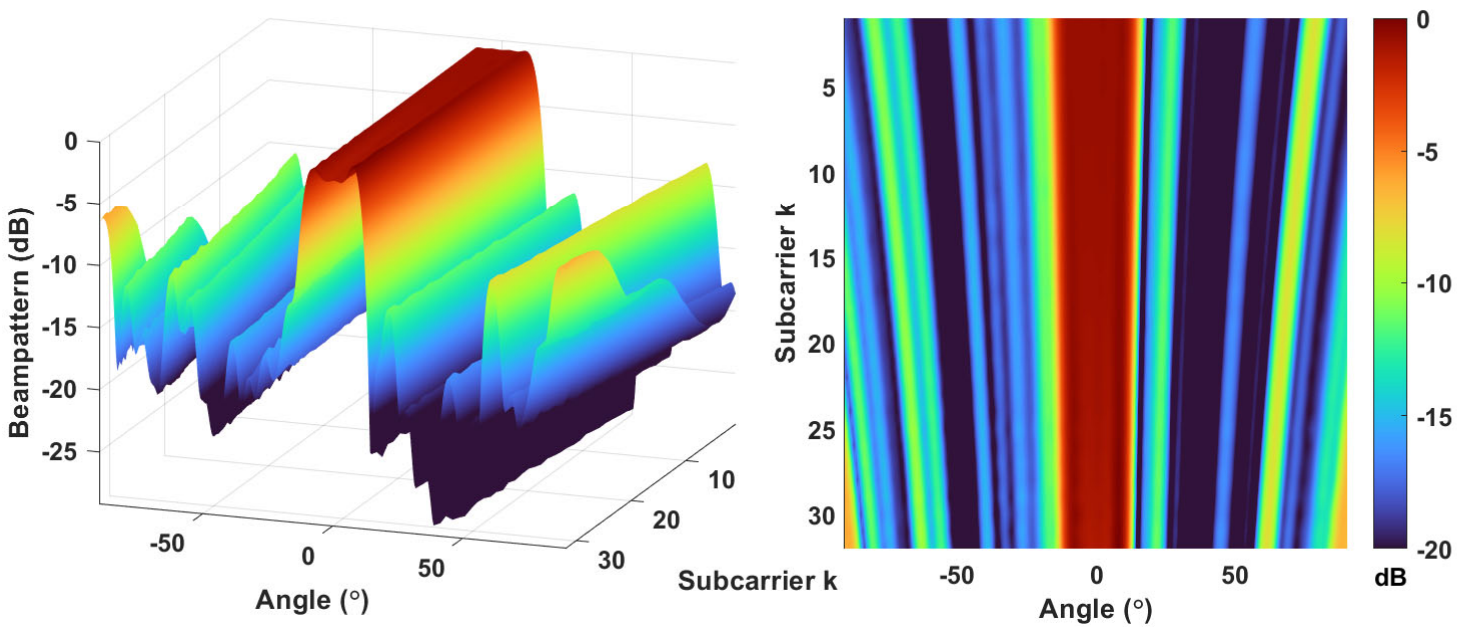}
\label{fig.2-3-5}
}\hspace{-0.9em}
\subfigure[]{
\includegraphics[width=0.327\linewidth]
{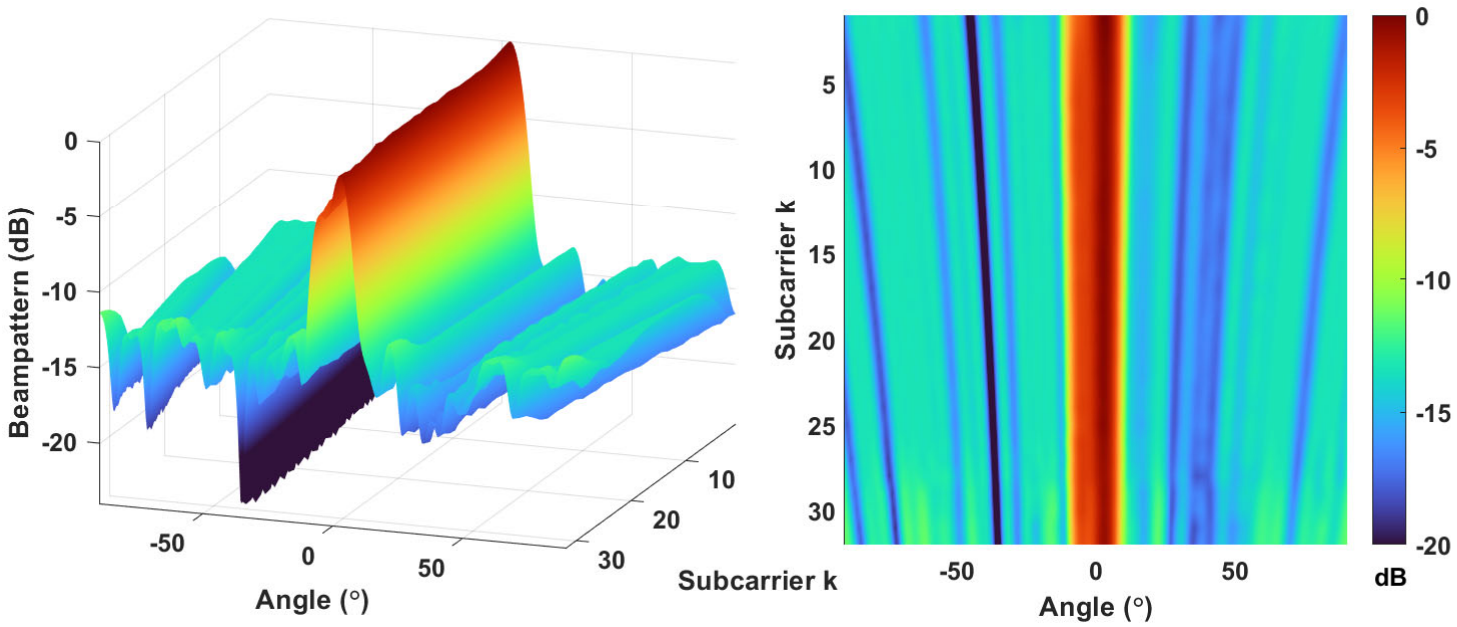}
\label{fig.2-3-4}
}\hspace{-0.9em}
\subfigure[]{
\includegraphics[width=0.327\linewidth]{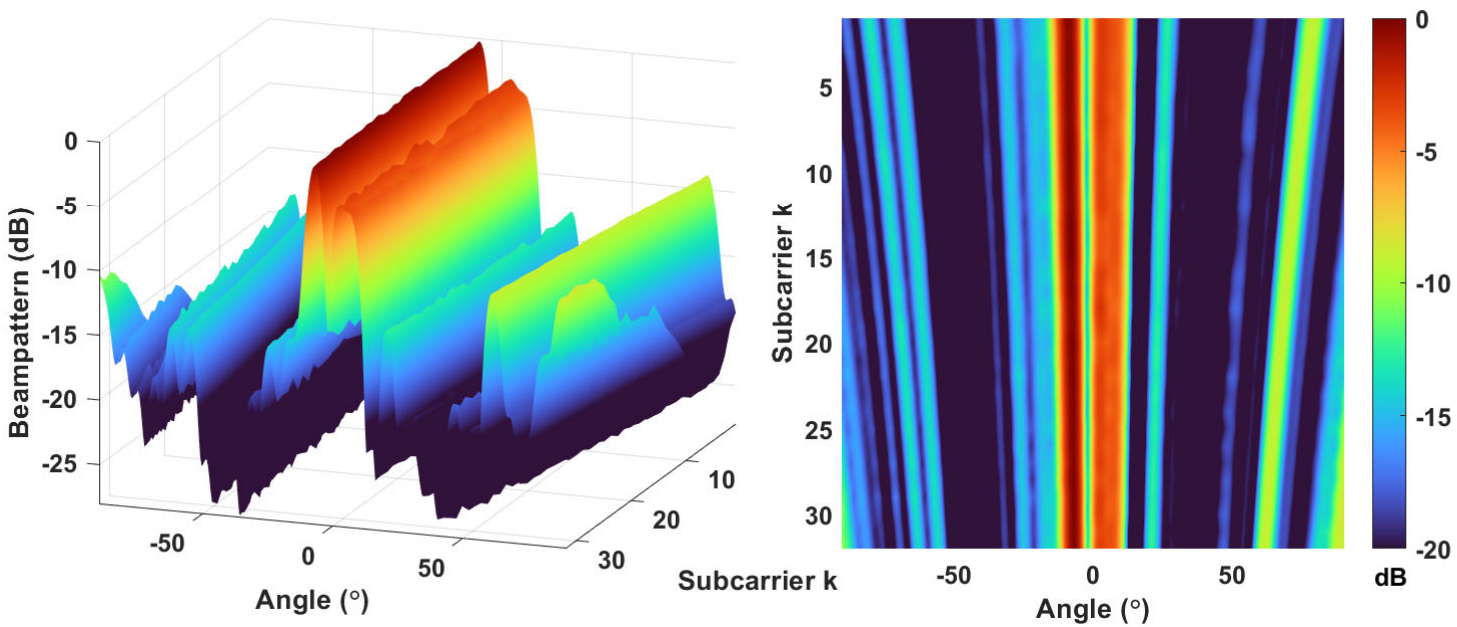}
\label{fig.2-3-6}
}\vspace{-0.8em}
\caption{ Comparisons of space-frequency behaviors of the wideband OFDM-DFRC systems by using different HBF methods and setting different expected beampattern, with rate threshold $\chi=2$. Desired Beampattern A: (a) Obtained by the proposed HBF for SD task. (b) Obtained by the proposed HBF for TT task. (c) Obtained by ISMR-based HBF. 
Desired Beampattern B: (d) Obtained by the proposed HBF for SD task. (e) Obtained by the proposed HBF for TT task. (f) Obtained by ISMR-based HBF.}
\label{fig.2-3}
    \vspace{-1em}
\end{figure*}

\begin{figure}[t]
    \centering
    \vspace{-0.5em}
    \includegraphics[width=0.95\linewidth]{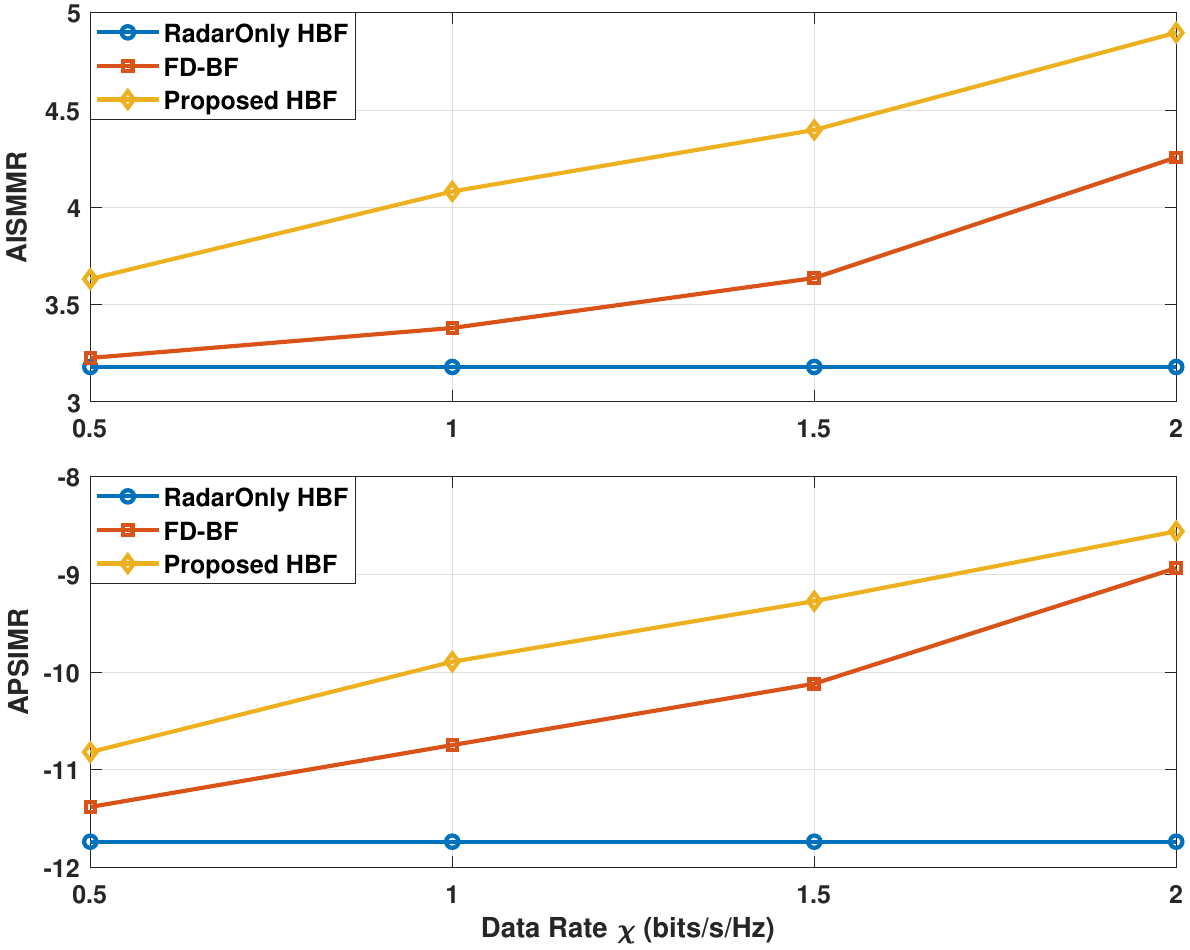}
    \vspace{-1em}
    \caption{{Comparisons of different methods: AISMMR and APSIMR versus the data rate $\chi$, with Desired Beampattern A.}}
    \label{fig.2-2}
\end{figure}

\subsubsection{Radar Space-Frequency Behaviors}

Fig. \ref{fig.2-3} demonstrates the space-frequency spectrum of the proposed task-oriented HBF and ISMR-based HBF with rate threshold $\chi =2$ and different desired radar beampatterns.
We can observe that the proposed task-oriented HBF for the SD task generates space-frequency spectra with a high mainlobe level on the whole mainlobe regions while the ISMR-based HBF fails to maintain all the levels of the mainlobe high.
In addition, the proposed task-oriented HBF for the TT task generates space-frequency spectra with low levels on the whole sidelobe region while the ISMR-based HBF brings relatively higher sidelobe levels in some directions.
This shows that the proposed task-oriented HBF can generate more excellent space-frequency spectra than the ISMR-based HBF in the SD task and TT task.
Therefore, it validates the superiority of the task-oriented HBF design for the OFDM-DFRC system with multi-task requirements.

Fig. \ref{fig.2-2} shows that the objective values become higher with the increase of the data rate threshold and compares the proposed HBF with other architectures.
It can be seen that the RadarOnly HBF achieves the lowest objective values for both scenarios, followed by the FD-BF.
As shown in the figure, the curves of the proposed HBF approach the curves of FD-BF.
This shows the effectiveness and superiority of the proposed HBF.

\subsubsection{Radar Detection Performance}

To further evaluate the radar detection performance of the proposed OFDM-DFRC system, assuming radar SINR is 18 dB, Fig. \ref{fig.2-4} presents the detection probability $P_{\rm d}$ versus the false alarm probability $P_{\rm fa}$.
It can be seen that for a fixed $P_{\rm fa}$, $P_{\rm d}$ of the SD task is higher than that of the TT task when the target is at $- 10 ^{\circ}$, and $P_{\rm d}$ of the TT task is higher than that of the SD task when the target is at $- 5 ^{\circ}$.
Although the detection probability of SD mode is low, it can have a rough detection among a wide target uncertainty region, which is more suitable for scanning.
Based on the prior detection information obtained, the TT mode can have a more precise detection among a narrow target uncertainty region.
Therefore, it validates that the proposed task-oriented HBF can satisfy different detection requirements for different task scenarios.

\begin{figure}[t]
    \centering
    \includegraphics[width=0.95\linewidth]{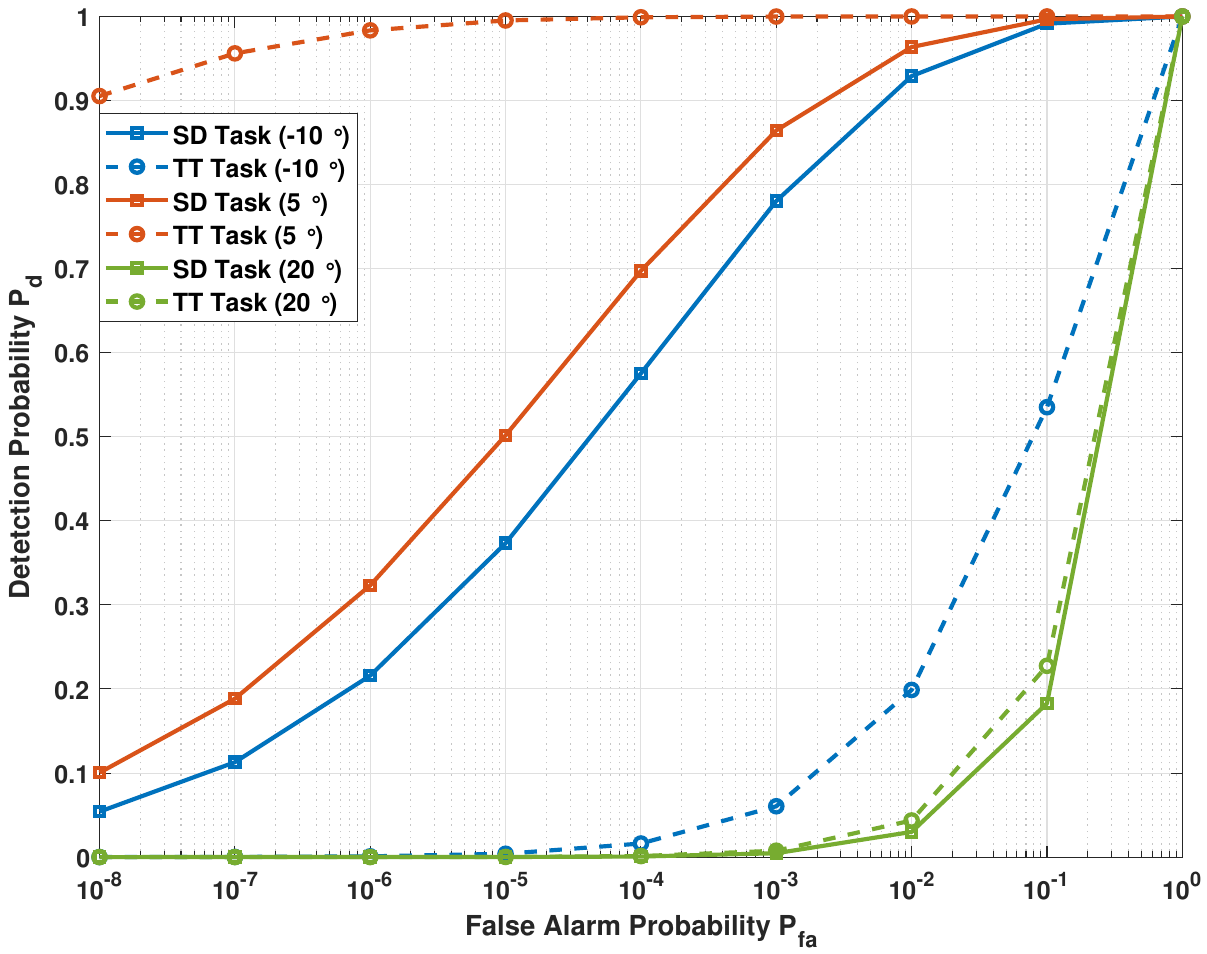}
    \vspace{-1em}
    \caption{Comparison of radar detection performance: the detection probability $P_{\rm d}$ versus false alarm probability $P_{\rm fa}$.}
    \label{fig.2-4}
    % \vspace{-1em}
\end{figure}

\vspace{-0.5em}
\subsection{ {Computation Time}}
\vspace{-0.3em}

\begin{table}[!h]
\vspace{-1.2em}
\begin{center}
    \caption{Average computation time of the proposed algorithm (per subcarrier, unit: second)}
    \vspace{-0.5em}
   \label{Table1}
\begin{tabular}{|c|c|c|c|c|}
\hline
($N_t$,$M_t$)      & (4,16) & (4,32) & (8,16) & (8,32) \\ \hline
Task 1 & 13.16  & 14.22  & 14.09  & 15.31  \\ \hline
Task 2 & 13.97  & 15.44  & 14.00  & 16.06  \\ \hline
\end{tabular}
\end{center}
\vspace{-1em}
\end{table}

In order to further evaluate the feasibility of the proposed algorithm in the multi-task OFDM-DFRC system, we set a different number of RF chains $N_t$ and transmit antennas $M_t$, applying Desired Beampattern B.
Note that all the numerical schemes are analyzed using Matlab 2022b version and performed in a standard PC with Intel(R) CPU(TM) Core i7-10700 2.9 GHz and 16 GB RAM.
Table \ref{Table1} shows the average computation time per subcarrier of the proposed method with a different number of $M_t$ and $N_t$. 
We can observe that a larger number of RF chains and transmit antennas prolong the computation time.
It indicates the feasibility and practicality of the proposed algorithm.

\vspace{-0.5em}
\section{Conclusion}
\vspace{-0.2em}
In this paper, we investigated a task-oriented HBF design for OFDM-DFRC systems.
We considered two task scenarios involving the radar scanning and detection task and the target tracking task.
To meet different tasks requirements, we introduced two novel radar beampattern metrics, the average integrated sidelobe to minimum mainlobe ratio (AISMMR) and the average peak sidelobe to integrated mainlobe ratio (APSIMR), to flexibly control the space-frequency spectra.
Besides, we set the data rate of each user on each subcarrier as the communication metric.
To devise beamforming schemes for different tasks, we formulated two HBF design problems by minimizing AISMMR and APSIMR respectively subject to the QoS constraint, power budget, and analog beamformer constraint.
To solve the high-dimensional and non-convex problems, we reformulated the problems and proposed a unified CADMM-based HBF algorithm to find their closed-form solutions.
The simulation results verify that the proposed task-oriented HBF can have flexible control of the space-frequency spectra by suppressing sidelobe levels and maintaining mainlobe levels in different radar sensing tasks.
Overall, the proposed task-oriented HBF design enhances the space-frequency behavior of the DFRC system in multiple tasks and guarantees the communication QoS simultaneously.
In future work, we can investigate an algorithm of lower complexity to further save computation time and achieve real-time computation.

\appendices{}

\section{Proof of Proposition 2}\label{app1}
With other variables being fixed, the original problem to update the analog beamformer ${{{\bf{F}}_{\text{RF}}}}$ is given by 
\begin{equation}
    \mathop {\min }\limits_{{{\bf{F}}_{\text{RF}}}}  {\| {{{\bf{V}}_{{k}}} - {{\bf{F}}_{\text{RF}}}{{\bf{F}}_k}} \|_F^2} \quad
    {\text{ s.t. }}| {{{\mathbf{F}}_{\text{RF}}}(i,j)} | = 1,{\text{ }}\forall  i, j \label{App_2} . 
\end{equation}
Obviously, the constant modulus constraint for the analog beamformer can be converted into an equivalent form:
\begin{equation}
    \left\{ {{\gamma _{i,j}}|{{\mathbf{F}}_{\text{RF}}}(i,j) = {e^{ - \jmath{\gamma _{i,j}}}},{\gamma _{i,j}} \in (0,2\pi ]} \right\}
    \label{App_3}
\end{equation}
Using the CCD method, we can optimize a single element in ${{{\bf{F}}_{\text{RF}}}}$, i.e., ${{{\bf{F}}_{\text{RF}}}}(i,j)$ by fixing other elements in ${{{\bf{F}}_{\text{RF}}}}$.
Then, aiming at optimizing ${{{\bf{F}}_{\text{RF}}}}(i,j)$, we can transform the original objective function as follows.
\begin{subequations}
\begin{align}
&\mathop {\min }\limits_{{{\bf{F}}_{\text{RF}}(i,j)}} {\| {{{\bf{V}}_{{k}}} - {{\bf{F}}_{\text{RF}}(i,j)}{{\bf{F}}_k}} \|_F^2}\\
=&\mathop {\min }\limits_{{{\bf{F}}_{\text{RF}}(i,j)}}   {\left \| {\bf{V}}_{k} \right \|}_F^2 -2\Re\{{{\bf{F}}_{\text{RF}}(i,j)} {{\bf{F}}_k}{\bf{V}}_{k}^H\} \\
&\qquad\;\;+ |{{\bf{F}}_{\text{RF}}(i,j)}|^2{\left \| {{\bf{F}}_k} \right \|}_F^2  \nonumber\\
\!\overset{(a)}{=}&\! \mathop {\min }\limits_{{{\bf{F}}_{\text{RF}}(i,j)}} \!  -2\Re\{{{\bf{F}}_{\text{RF}}(i,j)} {{\bf{F}}_k}{\bf{V}}_{k}^H\}\! +\! |{{\bf{F}}_{\text{RF}}(i,j)}|^2{\left \| {{\bf{F}}_k} \right \|}_F^2  \\
\overset{(b)}{=}&\mathop {\min }\limits_{{{\bf{F}}_{\text{RF}}(i,j)}} 2\Re\{{{\bf{F}}_{\text{RF}}(i,j)} ( {\bf{F}}_k(j,:){\bf{F}}_k^H{{\bf{T}}_{{\text{RF}},ij}^H(i,:)}   \\
&\qquad\;\;- {{\bf{F}}_k}(j,:){\bf{V}}_{k}^H(i,:) )\}\nonumber\\
\!=&\!\!\!\mathop {\min }\limits_{{{\bf{F}}_{\text{RF}}(i,j)}}\!\!\! 2\Re\{{{\bf{F}}_{\text{RF}}(i,\!j)}{\bf{F}}_k(j,\!:)\!( {\bf{F}}_k^H{{\bf{T}}_{{\text{RF}},ij}^H(i,\!:)}\! - \!{\bf{V}}_{k}^H(i,\!:) \!)\!\}
\end{align}
\end{subequations}
where Eq.(a) holds due to the constant transmit power constraint, Eq.(b) holds since each row only has one nonzero element for the analog beamformer and  ${\mathbf{T}}_{\text{RF},ij}^{}$ is ${{\mathbf{F}}_{\text{RF}}}$ whose $(i,j)$-th entry is zero.
Then, the original problem is converted into
\begin{subequations}
    \begin{align}
        \mathop {\min }\limits_{{{\mathbf{F}}_{\text{RF}}}(i,j)} \;\;&  2 \Re \left\{ {{{\mathbf{F}}_{\text{RF}}}(i,j){\psi _{i,j}}} \right\} + \text{const.}\\
        {\text{s.t. }} \quad & {{\mathbf{F}}_{\text{RF}}}(i,j) = {e^{ - \jmath{\gamma _{i,j}}}},{\gamma _{i,j}} \in (0,2\pi ],
    \end{align}
    \label{App_4}
\end{subequations} 
where ${\psi _{i,j}} = {{{\mathbf{F}}_k}(j,:)( {\mathbf{F}}_k^H {\mathbf{T}}_{\text{RF},ij}^H(i,:) - {{\mathbf{V}}_{k}^H(i,:) })}$.
Due to the relationship in \eqref{App_3}, the problem in \eqref{App_4} can be converted into a problem without constraints, which can be expressed as
\begin{equation}
    \mathop {\min }\limits_{{\gamma _{i,j}}} \quad\cos \left\{ { - {\gamma _{i,j}} + \angle {\psi _{i,j}}} \right\}
\end{equation}
Obviously, the optimal $\gamma _{i,j}^{\star} = \angle {\psi _{i,j}} + \pi$. Based on this, the optimal solution can be attained by substituting $\gamma _{i,j}^{\star}$. Finally, we get ${\mathbf{F}}_{\text{RF}}(i,j) = {\mathrm{exp} \{ - \jmath (\angle {\psi _{i,j}} + \pi)\}}$.
Thus, the proof is completed.
\ifCLASSOPTIONcaptionsoff
  \newpage
\fi

\footnotesize
\balance
 \bibliographystyle{IEEEtran}
 \bibliography{IEEEabrv,stan_ref}

\end{document}